\documentclass[twocolumn,trackchanges]{aastex63}

\usepackage{savesym}
\savesymbol{tablenum}

\usepackage{siunitx} % for "S" column type
\restoresymbol{SIX}{tablenum}

\usepackage{natbib}
\usepackage{amsbsy}

\usepackage{gensymb}
\usepackage[normalem]{ulem}
\DeclareSIUnit[space-before-unit=True]\erg{erg}
\DeclareSIUnit[space-before-unit=True]\ph{ph}
\DeclareSIUnit[space-before-unit=True]\pc{pc}
\DeclareSIUnit\mjd{MJD}
\usepackage[figuresright]{rotating}
\usepackage{textcomp}
\usepackage{hyperref}
\usepackage{eqnarray,amsmath}
\usepackage{graphicx}
\usepackage{multirow}

\received{Jun 7, 2022}
\revised{Dec 21, 2022}
\accepted{Jan 30, 2023}
\submitjournal{\apj}

\shorttitle{Gamma-Ray Upper Limits on SLSNe-I}
\begin{document}

% \title{Constraints on Late-Time Gamma-Ray Leakage from Superluminous Supernovae from VERITAS and {\it Fermi}-LAT}
\title{VERITAS and {\it Fermi}-LAT constraints on the Gamma-ray Emission from Superluminous Supernovae SN2015bn and SN2017egm}

\author[0000-0002-2028-9230]{A.~Acharyya}\affiliation{Department of Physics and Astronomy, University of Alabama, Tuscaloosa, AL 35487, USA}
\author[0000-0002-9021-6192]{C.~B.~Adams}\affiliation{Physics Department, Columbia University, New York, NY 10027, USA}
\author[0000-0002-3886-3739]{P.~Bangale}\affiliation{Department of Physics and Astronomy and the Bartol Research Institute, University of Delaware, Newark, DE 19716, USA}
\author[0000-0003-2098-170X]{W.~Benbow}\affiliation{Center for Astrophysics $|$ Harvard \& Smithsonian, Cambridge, MA 02138, USA}
\author[0000-0001-6391-9661]{J.~H.~Buckley}\affiliation{Department of Physics, Washington University, St. Louis, MO 63130, USA}
\author[0000-0002-8136-9461]{M.~Capasso}\affiliation{Department of Physics and Astronomy, Barnard College, Columbia University, NY 10027, USA}
\author[0000-0002-4661-7001]{V.~V.~Dwarkadas}\affiliation{Department of Astronomy and Astrophysics, University of Chicago, Chicago, IL, 60637}
\author[0000-0002-1853-863X]{M.~Errando}\affiliation{Department of Physics, Washington University, St. Louis, MO 63130, USA}
\author{A.~Falcone}\affiliation{Department of Astronomy and Astrophysics, 525 Davey Lab, Pennsylvania State University, University Park, PA 16802, USA}
\author[0000-0001-6674-4238]{Q.~Feng}\affiliation{Department of Physics and Astronomy, Barnard College, Columbia University, NY 10027, USA}
\author[0000-0002-8925-1046]{J.~P.~Finley}\affiliation{Department of Physics and Astronomy, Purdue University, West Lafayette, IN 47907, USA}
\author{G.~M.~Foote}\affiliation{Department of Physics and Astronomy and the Bartol Research Institute, University of Delaware, Newark, DE 19716, USA}
\author[0000-0002-1067-8558]{L.~Fortson}\affiliation{School of Physics and Astronomy, University of Minnesota, Minneapolis, MN 55455, USA}
\author[0000-0003-1614-1273]{A.~Furniss}\affiliation{Department of Physics, California State University - East Bay, Hayward, CA 94542, USA}
\author{G.~Gallagher}\affiliation{Department of Physics and Astronomy, Ball State University, Muncie, IN 47306, USA}
\author[0000-0001-7429-3828]{A.~Gent}\affiliation{School of Physics and Center for Relativistic Astrophysics, Georgia Institute of Technology, 837 State Street NW, Atlanta, GA 30332-0430}
\author[0000-0002-0109-4737]{W.~F~Hanlon}\affiliation{Center for Astrophysics $|$ Harvard \& Smithsonian, Cambridge, MA 02138, USA}
\author[0000-0003-3878-1677]{O.~Hervet}\affiliation{Santa Cruz Institute for Particle Physics and Department of Physics, University of California, Santa Cruz, CA 95064, USA}
\author{J.~Holder}\affiliation{Department of Physics and Astronomy and the Bartol Research Institute, University of Delaware, Newark, DE 19716, USA}
\author[0000-0002-1432-7771]{T.~B.~Humensky}\affiliation{Department of Physics, University of Maryland, College Park, MD 20742}
\author[0000-0002-1089-1754]{W.~Jin}\affiliation{Department of Physics and Astronomy, University of Alabama, Tuscaloosa, AL 35487, USA}
\author[0000-0002-3638-0637]{P.~Kaaret}\affiliation{Department of Physics and Astronomy, University of Iowa, Van Allen Hall, Iowa City, IA 52242, USA}
\author{M.~Kertzman}\affiliation{Department of Physics and Astronomy, DePauw University, Greencastle, IN 46135-0037, USA}
\author[0000-0003-4686-0922]{M.~Kherlakian}\affiliation{DESY, Platanenallee 6, 15738 Zeuthen, Germany}
\author[0000-0003-4785-0101]{D.~Kieda}\affiliation{Department of Physics and Astronomy, University of Utah, Salt Lake City, UT 84112, USA}
\author[0000-0002-4260-9186]{T.~K~Kleiner}\affiliation{DESY, Platanenallee 6, 15738 Zeuthen, Germany}
\author[0000-0002-5167-1221]{S.~Kumar}\affiliation{Physics Department, McGill University, Montreal, QC H3A 2T8, Canada}
\author[0000-0003-4641-4201]{M.~J.~Lang}\affiliation{School of Physics, National University of Ireland Galway, University Road, Galway, Ireland}
\author[0000-0003-3802-1619]{M.~Lundy}\affiliation{Physics Department, McGill University, Montreal, QC H3A 2T8, Canada}
\author[0000-0001-9868-4700]{G.~Maier}\affiliation{DESY, Platanenallee 6, 15738 Zeuthen, Germany}
\author[0000-0001-5544-1434]{C.~E~McGrath}\affiliation{School of Physics, University College Dublin, Belfield, Dublin 4, Ireland}
\author[0000-0002-2069-9838]{J.~Millis}\affiliation{Department of Physics and Astronomy, Ball State University, Muncie, IN 47306, USA and Department of Physics, Anderson University, 1100 East 5th Street, Anderson, IN 46012}
\author{P.~Moriarty}\affiliation{School of Physics, National University of Ireland Galway, University Road, Galway, Ireland}
\author[0000-0002-3223-0754]{R.~Mukherjee}\affiliation{Department of Physics and Astronomy, Barnard College, Columbia University, NY 10027, USA}
\author[0000-0002-8321-9168]{M.~Nievas-Rosillo}\affiliation{DESY, Platanenallee 6, 15738 Zeuthen, Germany}
\author[0000-0002-9296-2981]{S.~O'Brien}\affiliation{Physics Department, McGill University, Montreal, QC H3A 2T8, Canada}
\author[0000-0002-4837-5253]{R.~A.~Ong}\affiliation{Department of Physics and Astronomy, University of California, Los Angeles, CA 90095, USA}
\author[0000-0001-8965-7292]{S.~R.~Patel}\affiliation{DESY, Platanenallee 6, 15738 Zeuthen, Germany}
\author[0000-0002-7990-7179]{K.~Pfrang}\affiliation{DESY, Platanenallee 6, 15738 Zeuthen, Germany}
\author[0000-0001-7861-1707]{M.~Pohl}\affiliation{Institute of Physics and Astronomy, University of Potsdam, 14476 Potsdam-Golm, Germany and DESY, Platanenallee 6, 15738 Zeuthen, Germany}
\author[0000-0002-0529-1973]{E.~Pueschel}\affiliation{DESY, Platanenallee 6, 15738 Zeuthen, Germany}
\author[0000-0002-4855-2694]{J.~Quinn}\affiliation{School of Physics, University College Dublin, Belfield, Dublin 4, Ireland}
\author[0000-0002-5351-3323]{K.~Ragan}\affiliation{Physics Department, McGill University, Montreal, QC H3A 2T8, Canada}
\author{P.~T.~Reynolds}\affiliation{Department of Physical Sciences, Munster Technological University, Bishopstown, Cork, T12 P928, Ireland}
\author[0000-0002-7523-7366]{D. Ribeiro}\affiliation{Department of Physics, Columbia University, New York, NY 10027}
\author{E.~Roache}\affiliation{Center for Astrophysics $|$ Harvard \& Smithsonian, Cambridge, MA 02138, USA}
\author{J.~L.~Ryan}\affiliation{Department of Physics and Astronomy, University of California, Los Angeles, CA 90095, USA}
\author[0000-0003-1387-8915]{I.~Sadeh}\affiliation{DESY, Platanenallee 6, 15738 Zeuthen, Germany}
\author[0000-0001-7297-8217]{M.~Santander}\affiliation{Department of Physics and Astronomy, University of Alabama, Tuscaloosa, AL 35487, USA}
\author[0000-0003-1329-3909]{G.~H.~Sembroski}\affiliation{Department of Physics and Astronomy, Purdue University, West Lafayette, IN 47907, USA}
\author[0000-0002-9856-989X]{R.~Shang}\affiliation{Department of Physics and Astronomy, University of California, Los Angeles, CA 90095, USA}
\author[0000-0003-3407-9936]{M.~Splettstoesser}\affiliation{Santa Cruz Institute for Particle Physics and Department of Physics, University of California, Santa Cruz, CA 95064, USA}
\author[0000-0002-9852-2469]{D.~Tak}\affiliation{DESY, Platanenallee 6, 15738 Zeuthen, Germany}
\author{J.~V.~Tucci}\affiliation{Department of Physics, Indiana University-Purdue University Indianapolis, Indianapolis, IN 46202, USA}
\author[0000-0002-2126-2419]{A.~Weinstein}\affiliation{Department of Physics and Astronomy, Iowa State University, Ames, IA 50011, USA}
\author[0000-0003-2740-9714]{D.~A.~Williams}\affiliation{Santa Cruz Institute for Particle Physics and Department of Physics, University of California, Santa Cruz, CA 95064, USA}
\collaboration{56}{(VERITAS collaboration\footnote{\url{https://veritas.sao.arizona.edu}})}

\author[0000-0002-4670-7509]{B.~D.~Metzger}
\affiliation{Department of Physics, Columbia University, New York, NY 10027}
\affiliation{Center for Computational Astrophysics, Flatiron Institute, 162 5th Ave, New York, NY 10010, USA}

\author[0000-0002-2555-3192]{M.~Nicholl}
\affiliation{Birmingham Institute for Gravitational Wave Astronomy and School of Physics and Astronomy, University of Birmingham, Birmingham B15 2TT, UK}

\author[0000-0003-1336-4746]{I.~Vurm}
\affiliation{Tartu Observatory, University of Tartu, Toravere 61602, Tartumaa, Estonia}

\correspondingauthor{D. Ribeiro}
\email{d.ribeiro@columbia.edu}

% //
% V. V. Dwarkadas
% Department of Astronomy and Astrophysics, University of Chicago, Chicago, IL, 60637



\begin{abstract}
Superluminous supernovae (SLSNe) are a rare class of stellar explosions with luminosities ${\sim}$10--100 times greater than ordinary core-collapse supernovae. One popular model to explain the enhanced optical output of hydrogen-poor (Type I) SLSNe invokes energy injection from a rapidly spinning magnetar. A prediction in this case is that high-energy gamma rays, generated in the wind nebula of the magnetar, could escape through the expanding supernova ejecta at late times (months or more after optical peak). This paper presents a search for gamma-ray emission in the broad energy band from 100 MeV to 30 TeV from two Type I SLSNe, SN2015bn, and SN2017egm, using observations from {\it Fermi}-LAT and VERITAS. Although no gamma-ray emission was detected from either source, the derived upper limits approach the putative magnetar's spin-down luminosity. Prospects are explored for detecting very-high-energy (VHE; 100 GeV -- 100 TeV) emission from SLSNe-I with existing and planned facilities such as VERITAS and CTA. 

\end{abstract}


%% From the front matter, We move on to the body of the paper.

\section{Introduction}\label{sec:intro}

The recent growth of sensitive optical time-domain surveys has revealed and expanded exciting new classes of stellar explosions. These include superluminous supernovae, which can be up to 10--100 times more luminous than ordinary massive star explosions (e.g.~\citealt{Quimby2011,howell2013,inserra2013,nicholl2013,decia2018,lunnan2018,quimby2018}; see \citealt{Gal-Yam2019} for a recent review). Conventionally, the optical emission from most core-collapse supernovae is powered by the radioactive decay of $^{56}$Ni (Type Ib/c) and by thermal energy generated via shock heating of the stellar envelope (Type IIL, IIp). However, the peak luminosities of SLSNe greatly exceed the luminosity expected from those conventional mechanisms, and the origin of the energy is still debated. 

A popular model for powering the time-dependent emission of SLSNe, particularly the hydrogen-poor Type I class (SLSN-I), involves energy input from a young central engine, such as a black hole or neutron star, formed in the explosion. For example, the accretion onto the compact object from bound debris of the explosion could power an outflow which heats the supernova ejecta from within (\citealt{woosley2012,Quataert2012,margalit2016,moriya2018}). Alternatively, the central engine could be a strongly magnetized neutron star with a millisecond rotation period, whose rotationally powered wind provides a source of energetic particles that heat the supernova ejecta \citep{kasen2010,woosley2010,dessart2012a,metzger2015a,sukhbold2016}. The magnetar\footnote{To remain consistent with the SLSNe literature, the term magnetar is used throughout this paper. Magnetars generally have large dipole magnetic fields $B\simeq$ \SIrange{e13}{e15}{G} with a rotation period of a few seconds. In the case of the SLSN magnetar model, the radiation is extracted from the rotational energy of the young millisecond pulsars, but with large magnetic fields characteristic of magnetars.} model provides a good fit to the optical light curves of most SLSNe-I \citep{inserra2013,nicholl2017d}. Furthermore, analyses of the nebular spectra of hydrogen-poor SLSNe \citep{jerkstrand2017,nicholl2019} and Type Ib SNe \citep{milisavljevic2018} support the presence of a persistent central energy source, consistent with an energetic neutron star.

The details of how the magnetar would couple its energy to the ejecta are uncertain. Several models consider that the rotationally powered wind from a young pulsar inflates a nebula of relativistic electron/positron pairs and energetic radiation behind the expanding ejecta \citep{kotera2013,metzger2014b,murase2015}. At the wind-termination shock, the pairs are heated and radiate X-rays and gamma rays with high efficiency via synchrotron and inverse-Compton processes. Photons that evade absorption via $\gamma\textrm{-}\gamma$ pair creation in the nebula can be ``absorbed" by the ejecta further out, thermalizing their energy and directly powering the supernova's optical emission (e.g.~\citealt{metzger2014b, vurm2021}). 

Thermalization of the nebular radiation will be most efficient at early times, when the column through the ejecta shell and ``compactness'' of the nebula are at their highest. At these times one would expect the optical light curve to faithfully track the energy input of the central engine. However, as the ejecta expand, the radiation field dilutes and the shell becomes increasingly transparent to high-energy  and very-high-energy photons. The increasing transparency, and correspondingly decreasing thermalization efficiency, eventually causes the supernova's optical luminosity to drop below the rate of energy injection from the central engine \citep{chen2015,wang2015}, with the remaining radiation escaping directly from the nebula as gamma rays or X-rays (the putative ``missing" luminosity). 

\begin{figure}[ht]
    \centering
    \includegraphics[width=0.9\columnwidth]{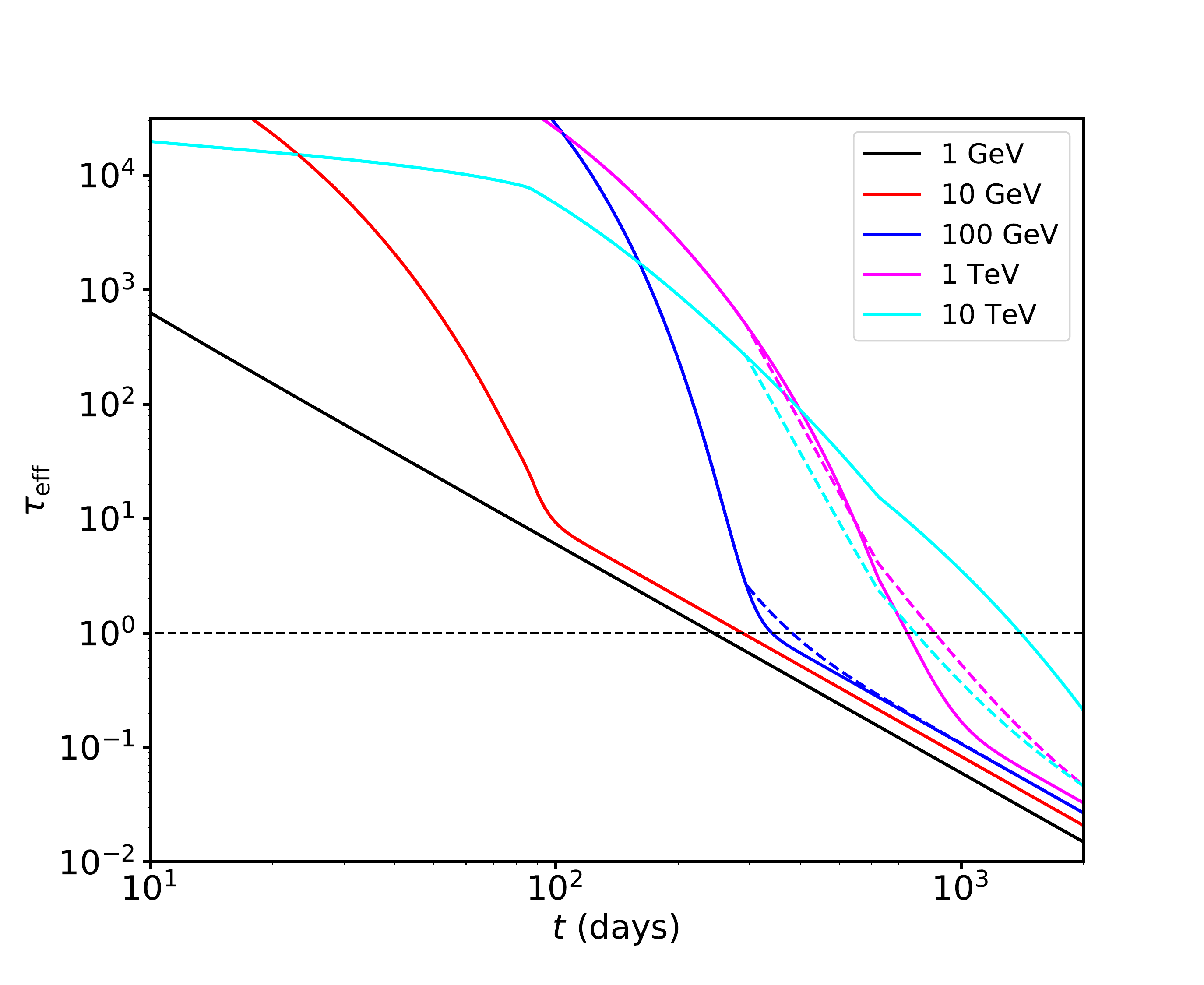}
    \includegraphics[width=0.9\columnwidth]{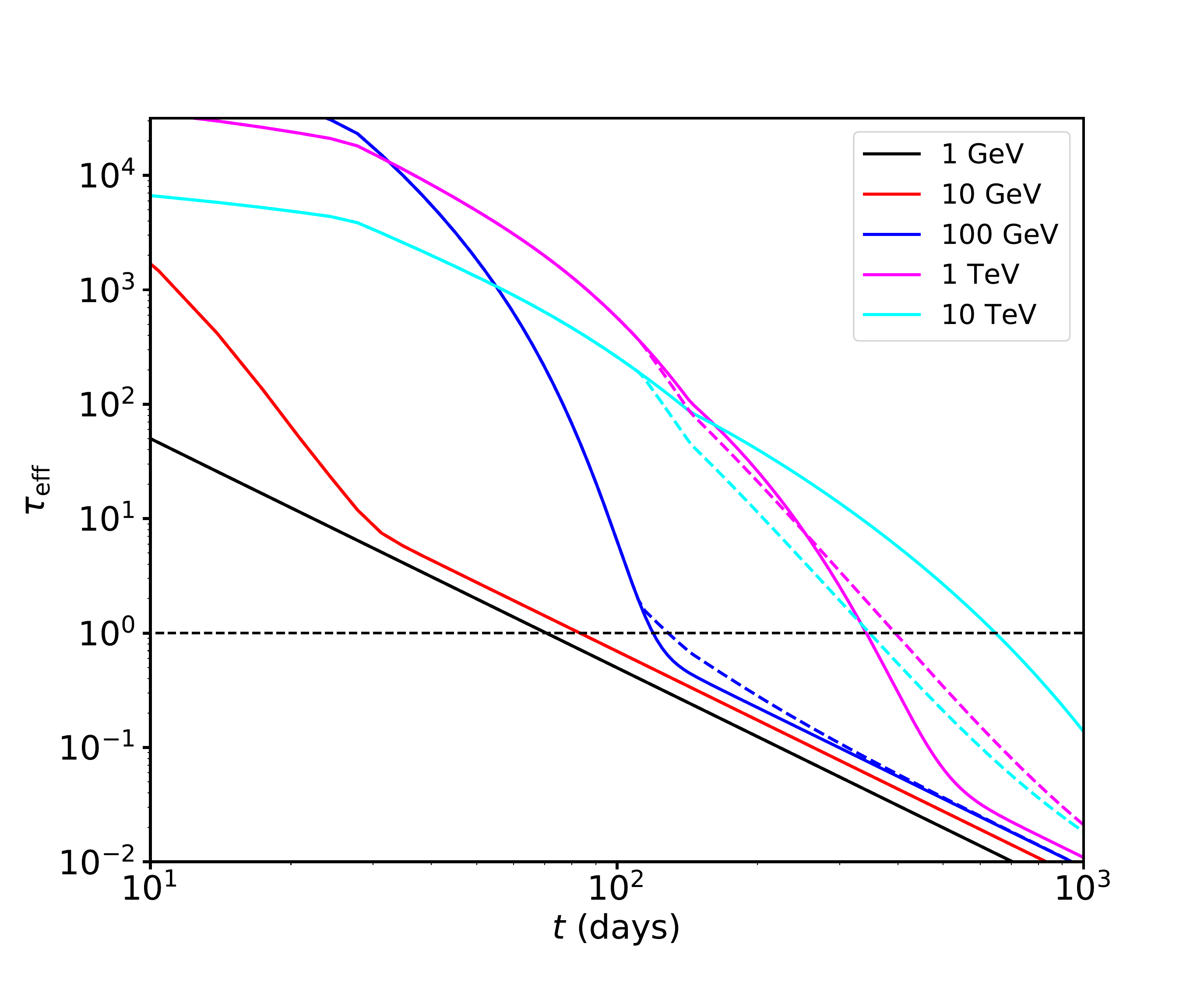}
    \caption{Optical depth at different photon energies as a function of time, calculated for ejecta properties (mass $M_{\odot}$, mean velocity, etc.) derived from observations of SN2015bn \citep{nicholl2018a} and SN2017egm \citep{nicholl2017c} shown in Table \ref{tab:event_physical_params}. Top: SN2015bn. Bottom: SN2017egm. The horizontal dotted line represents $\tau_{\rm eff} = 1$. The cross-sections for photon-photon and photon-matter pair production opacities are taken from \citet{zdziarski1989}. The solid lines correspond to target blackbody radiation temperature $T_{\rm eff} = (L_{\rm opt}/4\pi R^2)^{1/4}$, where $L_{\rm opt}$ and $R$ are the optical luminosity and ejecta radius, respectively. The dashed lines are computed with a temperature floor of $T = 4000$~K, to mimic the approximate spectrum in the nebular phase. Below ${\sim}10$~GeV the opacity is dominated by photon-matter pair production at all times. Above $100$~GeV, pair production on the thermal target radiation field dominates up to a few years.}
    \label{fig:tau}
\end{figure}

As the ejecta expand and the spin-down luminosity weakens, the conditions for various processes responsible for photon energy loss change and impact the effective optical depth. Within a few months, the effective optical depth to high-energy (HE; 100 MeV to 100 GeV) photons emitted from the central engine nears unity and, at several hundred days it reaches unity for very-high-energy (VHE; 100 GeV – 100 TeV) photons. 

Figure~\ref{fig:tau} shows examples of the effective optical depth through the ejecta for photons of various energies as a function of time. They have been calculated using time-dependent properties for the supernova ejecta and radiation field motivated by the observations of SN2015bn and SN2017egm, both particularly well-studied SLSNe-I explored in \cite{vurm2021}.

The dominant processes involved in the calculation of the gamma-ray optical depth include photon-matter and photon-photon interactions, particularly pair production on the nuclei and soft radiation fields in the ejecta. An accurate treatment considering the radiation transport is discussed in depth in \citet{vurm2021}. The standard version of the magnetar model \citep{kasen2010, woosley2010, nicholl2017d} does not consider this time-dependent calculation and relies on constant effective  opacities to optical and high-energy photons. Figure~\ref{fig:tau} provides a useful guiding timescale for when to consider gamma-ray emission at various energies, calculated with the model in \citet{vurm2021} using the ejecta properties fit to the optical data in Table \ref{tab:event_physical_params}.

Given its comparatively nearby distance at z=0.1136, SN2015bn is an excellent candidate event to test the magnetar hypothesis. The optical light curve shows a steepening from $\propto t^{-2}$ decay to $\propto t^{-4}$ around ${\sim}200$ days \citep{nicholl2018a}. This behavior is consistent with a leakage of high-energy radiation from a magnetar nebula \citep{nicholl2018a}. A deep search in the ${\sim}0.1-10$ keV X-ray band resulted in nondetections \citep{bhirombhakdi2018}, eliminating the possibility that leakage from the nebula occurs in the softer X-ray bands.

\citet{margutti2018b} present a similar search for late-time X-ray emission from a larger sample of SLSNe-I, mostly resulting in upper limits; however, see \citet{levan2013} for an X-ray detection of the SLSN-I SCP 06F6 that could still support the magnetar hypothesis. X-ray nondetections are not surprising, because the ejecta are likely to still be opaque in the $\lesssim 10$ keV band due to photoelectric absorption in the hydrogen-poor ejecta \citep{margalit2018a}. Intriguingly, \citet{Eftekhari2019} detected radio emission from the location of the SLSN PTF10hgi at 7.5 yr after the explosion and argued that the emission could be synchrotron emission from an engine-powered nebula.

Some effort has been underway to search for nebular leakage in the gamma-ray band. \citet{renault-tinacci2018} obtained upper limits on the $0.6-600$ GeV luminosities from SLSNe by a stacked analysis of 45 SLSNe with {\it Fermi}-LAT. The majority of their sample were SLSNe-I, the most likely class to be powered by a central engine; however the results were dominated by a single, extremely close Type II event (SLSN-II), CSS140222. Hydrogen-rich SLSNe make up the Type II class (SLSN-II), which are suggested to be powered by the interaction of the circumstellar medium with the supernova ejecta.  Nevertheless, even with CSS140222 included, the upper limits are at best marginally constraining on the inferred missing luminosity. 

In this paper, the search is expanded to gamma-ray emission from SLSNe-I in the HE to VHE bands using the {\it Fermi} Gamma-Ray Space Telescope and the ground-based VERITAS observatory. In particular, observations of SN2015bn and SN2017egm are presented here. SN2017egm is the closest SLSN-I to date in the Northern Hemisphere at z=0.0310 \citep{bose2017,nicholl2017c}. Observations of young supernovae with gamma-ray telescopes have been few, with no detections so far. Some tantalizing candidates like iPTF14hls and SN 2004dj have been explored with {\it Fermi}-LAT but are unconfirmed due to large localization regions overlapping with other gamma-ray candidates \citep{yuan2018,xi2020}. MAGIC carried out observations of a Type I SN \citep{ahnen2017a}. HESS observed a sample of core-collapse SNe \citep{Abdalla2019}, and later obtained upper limits on SN 1987A \citep{theh.e.s.s.collaboration2015}. Our observations are the first of superluminous supernovae. 

Throughout this paper, a flat $\Lambda$CDM cosmology is used, with $H_0 = \SI{67.7}{km.s^{-1}.Mpc^{-1}}$, $\Omega_{M}=0.307$, and $\Omega_{\Lambda} = 0.6911$ \citep{planckcollaboration2016}. The corresponding luminosity distances to SN2015bn and SN2017egm are  \SI{545.37}{\mega\pc} (z=0.1136) \citep{nicholl2016} and \SI{139.29}{\mega\pc} (z=0.0310) \citep{bose2017}.

\section{Observations and Methods} \label{sec:obs}
\label{sec:observations}
The superluminous supernovae SN2015bn and SN2017egm were observed with {\it Fermi}-LAT and VERITAS during 2015--2016 and 2017--2020, respectively. SN2015bn is a SLSN-I explosion from 23 Dec 2014 (MJD 57014) and it peaked optically on 19 Mar 2015 (MJD 57100) \cite{nicholl2016a}. SN2017egm is a SLSN-I explosion from 23 May 2017 (MJD 57896) and it peaked optically on 18 Jun 2017 (MJD 57922) \cite{bose2017}. Some properties of the SLSNe are given in Table \ref{tab:event_physical_params}. Details regarding the optical, {\it Fermi}-LAT and VERITAS observations and the data-analysis methods are below.

%%%%%%%%%%%%%%%%%%%%%%%%%%%%%%%%%%
% Magnetar Parameters
%%%%%%%%%%%%%%%%%%%%%%%%%%%%%%%%%%
\begin{table}[ht]
    \begin{center}
        \caption{Properties of the SLSNe considered in this paper. }
        \begin{tabular}{cl|rr}
        \multicolumn{1}{l}{Parameter} & {(}Unit{)}           & {SN2015bn} & {SN2017egm} \\ \hline
        R.A.                            & $^{\circ}$           & 173.4232 & 154.7734  \\
        decl                           & $^{\circ}$           & 0.725    & 46.454    \\
        z                             & -                    & 0.1136   & 0.0310    \\
        $t_{0}^{(a)}$                 & MJD                  & 57014    & 57896     \\
        $t_{pk}^{(b)}$                & MJD                  & 57100    & 57922     \\ \hline
        $P_0^{(c)}$                   & ms                   & $2.50^{+0.29}_{-0.17}$     & $5.83^{+0.73}_{-0.70}$      \\
        $B^{(d)}$                     & $10^{14}$ G          & $0.26^{+0.07}_{-0.05}$     & $0.94^{+0.13}_{-0.16}$      \\
        $M_{ej}^{(e)}$                & $M_{\odot}$          & $10.8^{+0.83}_{-1.34}$     & $2.99^{+0.30}_{-0.23}$      \\
        $\kappa^{(f)}$                & cm$^{2}$ g$^{-1}$     & $0.18^{+0.01}_{-0.02}$     & $0.12^{+0.04}_{-0.06}$      \\
        % $E_{SN}^{(g)}$                & $10^{51}$ erg        & 3.45     & 1.5      \\
        $v_{ej}^{(h)}$                & $10^{8}$ cm s$^{-1}$ & $5.68^{+0.16}_{-0.14}$     & $10.3^{+0.35}_{-0.27}$      \\
        $\kappa_{\gamma}^{(i)}$       & cm$^{2}$ g$^{-1}$     & $0.008^{+0.01}_{-0.01}$    & $0.080^{+0.15}_{-0.06}$     \\
        $M_{\rm NS}^{(j)}$            & $M_{\odot}$          & $1.84^{+0.28}_{-0.23}$     & $1.57^{+0.25}_{-0.29}$      
        \end{tabular}
        \label{tab:event_physical_params}        
    \end{center}
    Notes. The quantities $P_{0}$, $B$, $M_{\rm ej}$, $\kappa$, $E_{\rm SN}$, $v_{\rm ej}$, $\kappa_{\gamma}$ and $M_{\rm NS}$ were obtained from a best-fit to the UVOIR supernova light curves, with errors found in \cite{nicholl2017c, nicholl2017d}. 
    $^{(a)}$Epoch of explosion; $^{(b)}$Epoch of optical flux peak; $^{(c)}$Initial spin-period; $^{(d)}$Magnetic field strength of magnetar; $^{(e)}$Total mass, $^{(f)}$Effective opacity; $^{(g)}$Kinetic energy; $^{(h)}$Mean velocity of supernova ejecta; $^{(i)}$Gamma-ray effective opacity and $^{(j)}$Neutron star mass.
\end{table}

\subsection{{\it Fermi}-LAT} \label{subsec:Fermi}
The Large Area Telescope (LAT) on board the {\it Fermi} satellite has operated since 2008 \citep{Atwood2009}. It is sensitive to photons between \SI{{\sim}20}{\MeV} and \SI{{\sim}300}{\GeV} and has an ${\sim}60\degree$ field of view, enabling it to survey the entire sky in about 3 hours. 

The data were analyzed using the publicly available {\it Fermi}-LAT data with the \texttt{Fermitools} suite of tools provided by the {\it Fermi} Science Support Center (FSSC). Using the \texttt{Fermipy} analysis package \citep{wood2017}
\footnote{\url{https://fermipy.readthedocs.io/en/latest/} ; v0.19.0}, the data were prepared for a binned likelihood analysis in which a spatial spectral model is fit over the energy bins. The data were selected using the SOURCE class of events, which are optimized for point-source analysis, within a region of $15\degree$ radius from the analysis target position. Due to the effect of the Earth, a $90\degree$ zenith angle cut was applied to remove any external background events. The standard background models were applied to the test model, incorporating an isotropic background and a galactic diffuse emission model without any modifications. The standard 4FGL catalog was then queried for sources within the field of view and their default model parameters \cite{abdollahi2020}. 
 
Additional putative point sources were added to each field of view as needed to support convergence of the fit. These sources were added for all analysis time scales. This process continued until the distribution of test statistics for the field of view was Gaussian with standard deviation near 1 and mean centered at 0, and the residual maps were near uniformly 0 without strong features. These conditions indicate the appropriate coverage of spectral sources within the analysis was reached and no putative sources are missing. The fitting process is performed in discrete energy bins while optimizing the spectral shape, but the distribution of test statistics is evaluated with the stacked data spanning the full energy range. With the improvements to {\it Fermi}-LAT low-energy sensitivity in PASS8 reconstruction, the low-energy bin covering $100-612$ MeV was also added. 
 
In the case of both SN2015bn and SN2017egm, the data were fitted with a power-law spectral model, $N(E) = N_{0} E^{\Gamma}$, with a free prefactor and a fixed photon index $\Gamma$ of -2.0.  From the fit, the reported flux upper limit was found using a 95\% confidence level with the bounded Rolke method \citep{rolke2005}. In all cases reported here, the upper limit reported is the integral energy flux, integrated over the energy ranges described for each case, which has units of \si{MeV cm^{-2}.s^{-1}}. This flux is converted to luminosity with the adopted distance for each event.

SN2015bn was observed from 2014 December 23 to 2018 Mar 23. This observation period begins after the explosion, and is binned in a few windows to account for the absorption of low-energy gamma rays by the ejecta at early times (Figure~\ref{fig:tau}). The first $\sim90$ days is observed to make sure there are no early emission during the expected absorption period. The data were thereafter binned in time intervals of six months to maximize observation depth and sensitivity to time-dependent variation. SN2017egm was observed 2017 May 23 to 2020 Aug 21. Again, this period covers the 3.5 yr from the discovery date, starting with $\sim90$ days after the explosion and split into six 6 month bins thereafter. The 3.5 yr observation period is selected to cover approximately 1000 days after the explosion. After this period, it is expected that the predicted luminosity will have decreased below the {\it Fermi}-LAT detectable limit. 

SN2015bn is within 5\textdegree\ of the Sun each year in August, so a one-month time cut is applied to each relevant time bin (to cover an $\sim15$\textdegree\ radius field of view). SN2017egm is not near the the path of the Sun, so this cut was not applied.

\subsection{VERITAS} \label{subsec:veritas}
The Very Energetic Radiation Imaging Telescope Array System (VERITAS) is an imaging atmospheric cherenkov telescope (IACT) array at the Fred Lawrence Whipple Observatory (FLWO) in southern Arizona, USA \citep{weekes2002, Holder2006}. It consists of four 12 m telescopes separated by approximately $100$ m, and the observatory is sensitive to photons within the energy range $\backsim$\SI{100}{GeV} to $\backsim$\SI{30}{TeV}. The instrument has an angular resolution (68\% containment) of $\backsim$0.1\degr\ at \SI{1}{TeV}, an energy resolution of $\backsim$15\% at \SI{1}{TeV}, and 3.5\degr\ field of view.

VERITAS serendipitously observed SN2015bn for a total of 1.01 hr between 2015 May 7 and 2015 May 22, approximately 135 days from explosion (49 days from the date of peak magnitude), as a part of an unrelated campaign. Another 1.7 hr were taken between 2016 May 25 and 30. Data were taken in good weather and dark sky conditions. Since SN2015bn was not the target source, its sky position averages 1.4\textdegree~from the center of the camera.  

VERITAS directly observed SN2017egm for 8.7 hr between 2019 Mar 24 and 2019 Apr 5, under dark sky conditions, as part of a Directors Discretionary Time (DDT) campaign, approximately 670 days from explosion. This target was triggered based on the predicted gamma-ray luminosity (see section \ref{sec:MagnetarSpinDown} and appendix \ref{sec:appendix} for a description) derived from the optical observation. Although it was almost two years after the explosion, the nearby distance yielded a gamma-ray luminosity prediction still within reach of VERITAS, making this an enticing target to follow up.

The SN2017egm data in this paper were taken using ``wobble" pointing mode, where the source is offset from the center of the camera by $0.5\degree$. This mode creates space for a radially symmetric off region to be used for background estimation in the same field of view, saving time from targeted background observations that contain the same data observing conditions. The data were processed with standard VERITAS calibration and reconstruction pipelines, and then cross-checked with a separate analysis chain \citep{cogan2008,Maier2017}. 
 
Using an Image Template Method (ITM) to improve event angular and energy reconstruction \citep{christiansen2017}, analysis cuts are determined with a set of a priori data-selection cuts optimized on sources with a moderate power-law index (from -2.5 to -3).

Unfortunately, the large offset on SN2015bn due to the serendipitous observation precludes us from using ITM in the analysis, so in that case SN2015bn is analyzed without templates by calculating image moments directly from candidate images triggered by the camera \citep{cogan2008,Maier2017}.   In both cases, the signal and background counts are determined using the reflected region method.

The upper limit is calculated for both SN2015bn and SN2017egm. The bounded Rolke method for upper limit calculation is used, assuming a power law spectrum with index of -2.0 and 95\% confidence level \citep{rolke2005}. Since the calculation of the upper limit depends on the underlying spectral model, a range of power-law spectral indices from -2 to -3 was computed to estimate impact of the model dependence. In all cases reported here, the upper limit reported is the integral photon flux, integrated over the energy ranges described for each case, which has units of \si{cm^{-2}.s^{-1}}. This flux is converted to integral energy flux using the same spectral model so that the luminosity can be computed with the adopted distance.

\begin{figure*}[ht]
    \centering
    \includegraphics{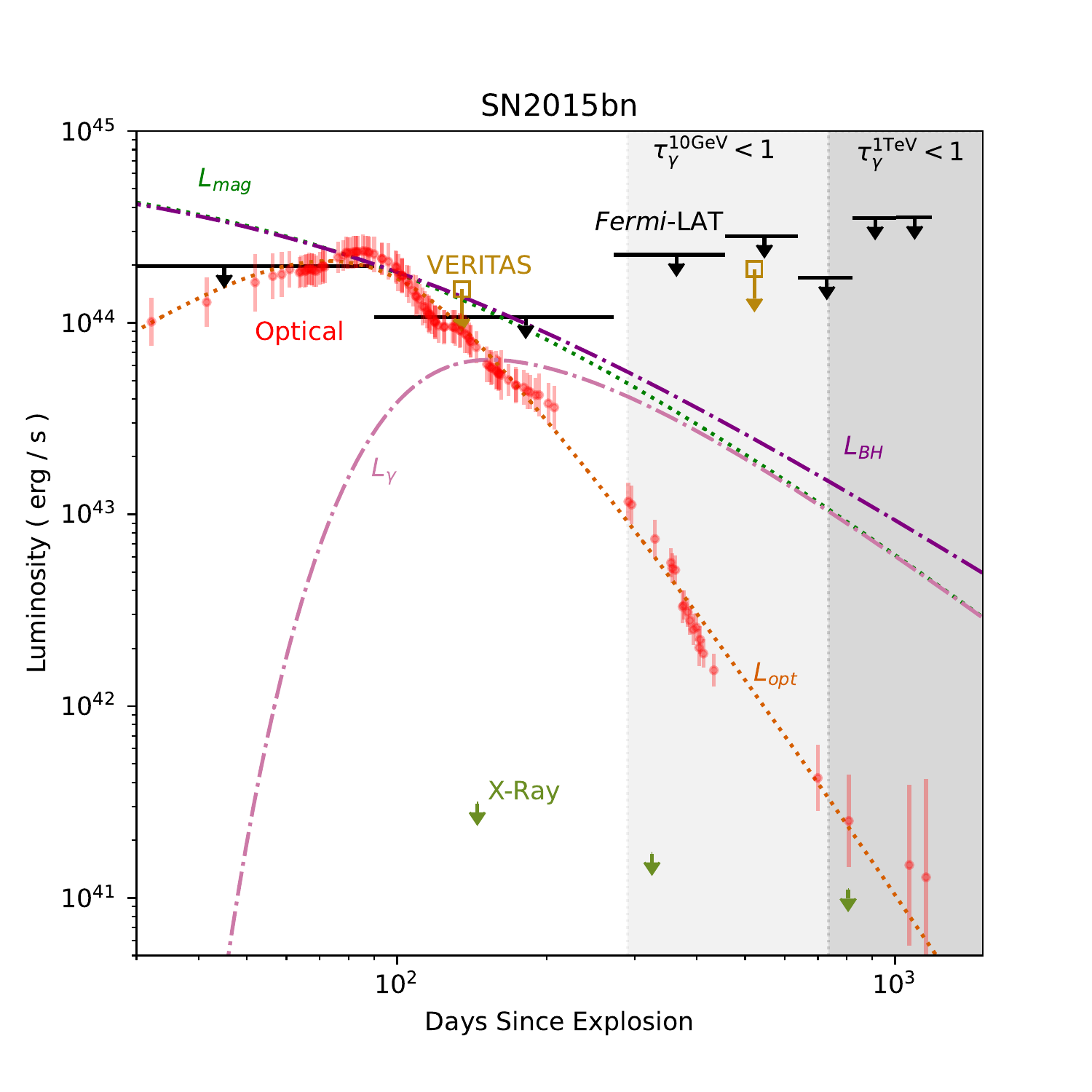}
    \caption{Light curves of SN2015bn spanning 30-1500 days after explosion. Curves shown include (1) the (thermal) supernova luminosity, $L_{\rm opt}$, fit to UVOIR bolometric luminosity data (in red; \citealt{nicholl2018a}) to obtain the magnetar parameters; (2) magnetar spin-down luminosity, $L_{\rm mag}$ (green dotted lined); and (3) predicted gamma-ray luminosity that escape the ejecta, $L_{\gamma}$ (pink dotted-dashed line; Equations \ref{eq:Lmag}, \ref{eq:trapped} and \ref{eq:leaking}). Black bars show {\it Fermi}-LAT upper limits reported for six 180 day bins starting ${\sim}90$ days after explosion. The olive open box shows the VERITAS integral energy flux} upper limit taken ${\sim}135$ days after the explosion, with EBL absorption correction applied. Upper limits on the 0.2-10 keV X-ray luminosity from {\it Chandra} are from \citet{bhirombhakdi2018} in green. Gray shaded regions labeled ``$\tau_{\gamma} <1$" show the approximate time after which gamma rays of the indicated energy should escape ejecta, based on Figure \ref{fig:tau}. A purple dotted-dashed line shows the engine luminosity, $L_{\mathrm{BH}}$ (Eq.~\ref{eq:L_BH}), in an alternative model in which the supernova optical luminosity is powered by fallback accretion onto a black hole. All upper limits denote the 95\% confidence level.
    \label{fig:SN2015bn_lc}
\end{figure*}

\begin{figure*}[ht]
    \centering
    \includegraphics{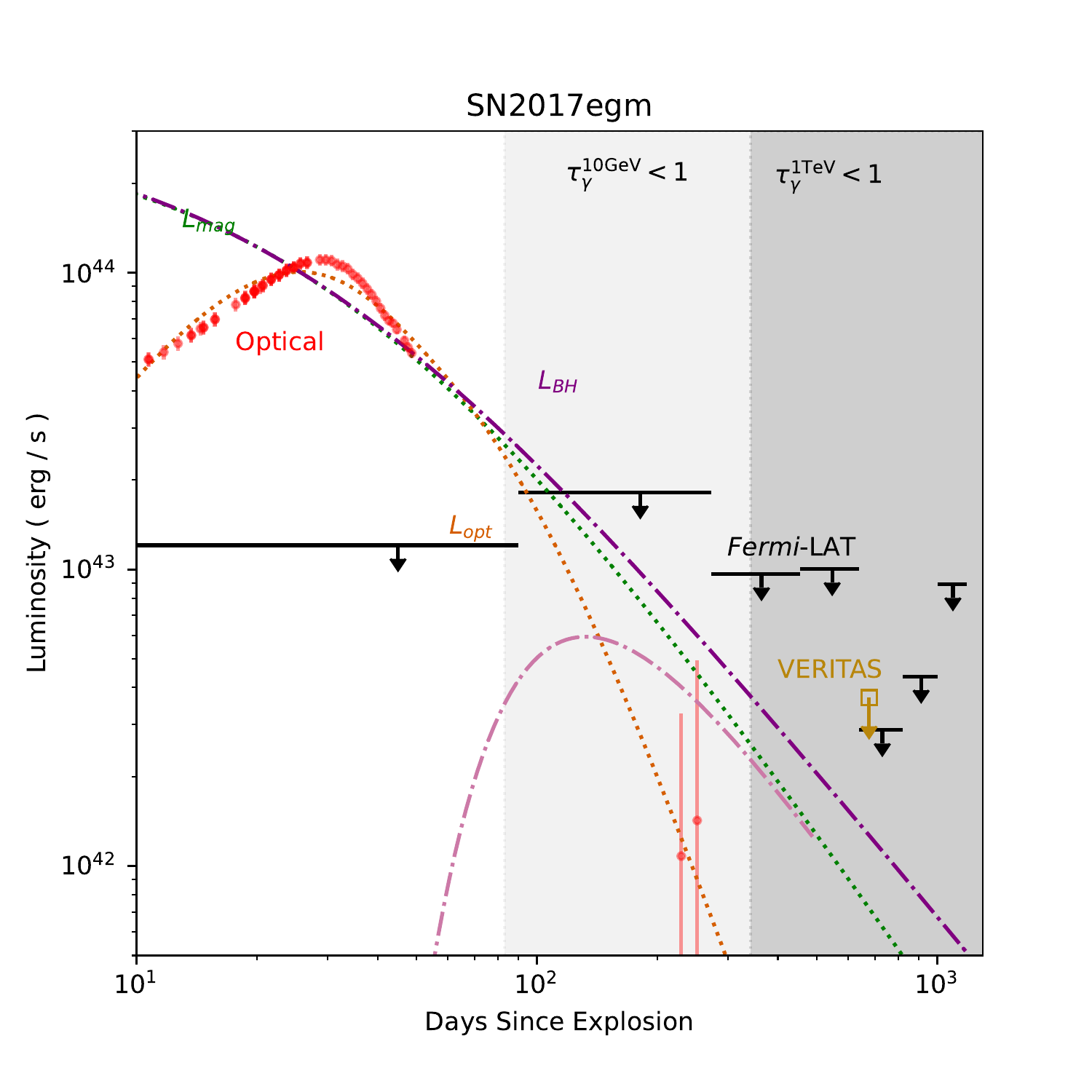}
    \caption{The SN2017egm light curve spanning 10-1300 days after explosion, following the same format as Figure~\ref{fig:SN2015bn_lc}. UVOIR data are shown in red \citep{bose2017, nicholl2017c}. Integral energy flux upper limits from {\it Fermi}-LAT are reported for six 180 day bins starting ${\sim}90$ days after the explosion. Integral energy flux upper limits are shown for VERITAS data taken ${\sim}670$ days after explosion, with EBL absorption correction applied. The maximum luminosity of the black hole accretion model $L_{\mathrm{BH}}$ (Eq. \ref{eq:L_BH}) is shown in purple. All upper limits denote the 95\% confidence level.}
    \label{fig:SN2017egm_lc}
\end{figure*}

\begin{table*}[ht]
    \begin{center}
        \caption{Results from VERITAS observations for both epochs of SN2015bn, and SN2017egm. }
        \label{tab:veritas_total_results}
        \begin{tabular}{cl|rrr}
        \multicolumn{1}{l}{Parameter} & {(}Unit{)} & SN2015bn$_{1}$ & SN2015bn$_{2}$ & SN2017egm \\ 
        \hline
        \hline
        Start (MJD) & [day] & 57149 & 57533 & 58566 \\
        End (MJD) & [day] & 57164 & 57538 & 58578 \\
        Live time & [hr] & 1.0 & 1.8 & 8.7 \\
        On & [event] & 4 & 10 & 49\\
        Off & [event] & 179 & 188 & 596 \\
        $\alpha^{(a)}$ & - & 0.0286 & 0.0299 & 0.0634 \\
        Excess & [event] & -1.1 & 4.4 & 11.2 \\
        Significance & [$\sigma$] & -0.5 & 1.7 & 1.6 \\
        Flux UL & [$\SI{e-13}{cm^{-2}~s^{-1}}$] & 28.5 & 27.8 & 10.2 \\
        $E_{\rm  threshold}$ & [GeV] & $>320$ & $>420$ & $>350$ \\ 
        \end{tabular}
    \end{center}
    Notes. Shown are the quality selected livetime, number of gamma-ray-like events in the on and off-source regions, the normalization, the observed excess of the gamma-rays and the statistical significance. The integral flux upper limit is shown for the given energy threshold, without EBL absorption correction, integrated up to \SI{30}{\TeV}.
    $^{(a)}$ Ratio of relative exposure for on and off regions.
\end{table*}

\section{Results}
\label{sec:results}

No statistically significant detections were made of either SN2015bn or SN2017egm across the energy range 100 MeV to 30 TeV. Integral energy upper limits are reported for the energy ranges given for each instrument. Figure~\ref{fig:SN2015bn_lc} and \ref{fig:SN2017egm_lc} show the {\it Fermi}-LAT and VERITAS upper limits in comparison to the supernova optical light curves and the theoretically predicted escaping luminosity from the magnetar model. 

\subsection{Optical}\label{sec:optical_result}
The SN2015bn integrated ultraviolet-optical-infrared (UVOIR) light-curve data are reproduced here from previous analyses \citep{nicholl2016, nicholl2016a, nicholl2018a}. To produce these bolometric light curves, the multiband optical data were interpolated and integrated at each epoch using the code \texttt{superbol} \citep{nicholl2018}. 

Similarly, the SN2017egm UVOIR data are also reproduced here with \texttt{superbol} \citep{bose2017, nicholl2017c}.

\subsection{{\it Fermi}-LAT}
Both SN2015bn and SN2017egm are not statistically significant sources in the first $\sim90$ days or the subsequent 6 month bin starting 90 days after the explosion. These sources also remain undetected in any of the following 6 month bins, and in the multiyear data sets. 

The evaluation of the integral energy flux upper limit for the {\it Fermi}-LAT observations within each time bin was performed assuming a power-law spectral model with an index of -2. The model dependence of this calculation naturally impacts the interpretations in section \ref{sec:discussion}, so the the fit was performed with indices 2, 2.5 and 3 to find the impact of the model on the final upper limit. An uncertainty of about $10\%$ was found based on varying the index.

SN2015bn is found to have test statistic (TS) of 0.06,
% (about 0.24 sigma)
with 12 predicted events above the isotropic diffuse background $\simeq \num{4.8e4} $ events over the entire period. The flux upper limit is $\SI{1.6e-6}{\MeV.\cm^{-2}.\s^{-1}}$ over the energy range \SI{100}{\MeV} to \SI{500}{\GeV}. In the first $\sim90$ days after the explosion, where the gamma ray emission is not expected due to the high gamma-ray absorption (see Figure \ref{fig:tau}), the flux upper limit is $\SI{3.5e-6}{\MeV.\cm^{-2}.\s^{-1}}$ over the energy range \SI{100}{\MeV} to \SI{500}{\GeV}, with a TS of 0. For the first 6-month period, when the signal is most likely, the flux upper limit is $\SI{1.9e-6}{\MeV.\cm^{-2}.\s^{-1}}$ for $TS \simeq 0$, consistent with a nondetection. All of the following 6 month bins reported nondetections with $TS<2$. 

SN2017egm is found to have $TS=4.4$,
% (about 2.1 sigma)
with 43 predicted events above the isotropic diffuse background $\simeq \num{5.9e4} $ events. The flux upper limit is $\SI{1.2e-6}{\MeV.\cm^{-2}.\s^{-1}}$ over the energy range \SI{100}{\MeV} to \SI{500}{\GeV}. In the first $\sim90$ days after the explosion, where the gamma ray emission is not expected due to the high gamma-ray absorption (see Figure \ref{fig:tau}), the flux upper limit is $\SI{3.2e-6}{\MeV.\cm^{-2}.\s^{-1}}$ over the energy range \SI{100}{\MeV} to \SI{500}{\GeV}, with a TS of 0. For the first 6-month period, when the signal is most likely, the flux upper limit is $\SI{4.9e-6}{\MeV.\cm^{-2}.\s^{-1}}$ for $TS=\num{10.1}$, consistent with a non-detection. All of the following 6-month bins reported non-detections with $TS<1$.

\subsection{VERITAS}
Table \ref{tab:veritas_total_results} reports the results from VERITAS observations of SN2015bn and SN2017egm. Each observation is consistent with a nondetection. The significance of each excess of observed events above background is below 2 standard deviations ($\sigma$). The flux upper limits are also given, calculated by integrating above the threshold energy of the instrument.

The statistical significance of an excess is estimated using Equation 17 of Li \& Ma \citep{li1983}. SN2015bn has significance value of $-0.5 \sigma$ in the first epoch observation. The integral flux upper limit from \SIrange{0.32}{30}{\TeV} for SN2015bn is \SI{2.85e-12}{cm^{-2}.s^{-1}}, which corresponds to an upper limit on the luminosity of \SI{1.27e44}{\erg.s^{-1}} at a redshift of 0.1136.  Due to the serendipitous nature of the observation, SN2015bn is significantly off axis, which lowers the instrument sensitivity at the energy threshold of \SI{320}{\GeV}.  Additionally, a 10\% systematic uncertainty is added to the flux normalization and reported energy threshold due to instrument degradation during the period of 2012-2015 \cite{nievasrosillo2021}. This uncertainty is derived empirically from the observation of the Crab Nebula over the same period. During the second observation in 2016, SN2015bn was found to have a significance of 1.7. The integral flux upper limit from \SIrange{0.42}{30}{\TeV} for SN2015bn is \SI{2.78e-12}{cm^{-2}.s^{-1}}, which corresponds to an upper limit on the luminosity of \SI{1.60e44}{\erg.s^{-1}}. 

For SN2017egm, the Li \& Ma significance value is $0.2 \sigma$ and an integral upper limit from \SIrange{0.35}{30}{\TeV} is \SI{1.0238e-12}{cm^{-2}.s^{-1}}, which corresponds to an upper limit on the luminosity of \SI{3.54e42}{\erg. s^{-1}} above the energy threshold of \SI{350}{\GeV} at redshift z=0.0310. The systematic correction due to instrument degradation during the period of 2012-2019 is applied automatically with the use of the throughput-calibrated analysis templates \citep{nievasrosillo2021}. In the cases of both SN2015bn and SN2017egm, the impact of varying the power-law model index parameter from -2 to -5 is about 10\%, which is a negligible in the context of their respective light curves.

VHE photons are absorbed by the extragalactic background light (EBL) throughout the universe, so the flux must be corrected to account for the missing photons. This absorption is energy and redshift dependent. Deabsorption is applied to the flux using the model of \citet{Dominguez2011}.  
The EBL deabsorption factor was convolved with the upper limit calculation, assuming the same spectral shape (a power law with the photon index of -2.0). 
The deabsorbed integral photon upper limit for SN2015bn within the energy range \SIrange{0.32}{30}{\TeV}, is \SI{3.36e-12}{cm^{-2}.s^{-1}}, which corresponds to a luminosity upper limit of \SI{1.49e44}{\erg.s^{-1}}. For the second observation, the deabsorbed integral photon upper limit for SN2015bn within the energy range \SIrange{0.42}{30}{\TeV}, is \SI{3.30e-12}{cm^{-2}.s^{-1}}, which corresponds to a luminosity upper limit of \SI{1.91e44}{\erg.s^{-1}}. For SN2017egm, with a slightly smaller energy range \SIrange{0.350}{30}{\TeV}, the deabsorbed integral photon flux is \SI{1.07e-12}{cm^{-2}.s^{-1}}, which corresponds to a luminosity upper limit of \SI{3.70e42}{\erg.s^{-1}}. These EBL-corrected values are plotted in Figure \ref{fig:SN2015bn_lc} and Figure \ref{fig:SN2017egm_lc}.

\section{Discussion}
\label{sec:discussion}
The source of the extra luminosity powering SLSNe-I may be found in the signature of its late-time gamma-ray emission. This section explores the HE to VHE emission hundreds of days after the explosion. The following models with a gamma-ray emission component for the powering mechanism are discussed: (1) magnetar central engine (see section \ref{sec:MagnetarSpinDown}), (2) black hole central engine (see section \ref{sec:BlackHole}), and (3) circumstellar interaction (see section \ref{sec:Circumstellar}).

\subsection{Magnetar Central Engine} \label{sec:MagnetarSpinDown}

The most promising mechanism for powering SLSNe-I is the rotational energy input from a central magnetar. In this scenario, a young pulsar or magnetar inflates a nebula of relativistic particles, which radiate high-energy gamma rays and X-rays. This section initially explores a simple implementation of the magnetar model (see Appendix \ref{sec:appendix} for full description), followed by a more complete model described in detail in \citet{vurm2021} for both SN2015bn and SN2017egm. The application of this so-called self-consistent model is necessary to directly predict the energy-dependent luminosities within the energy ranges of the {\it Fermi}-LAT and VERITAS observations, a major contribution that is not possible with simpler implementation described in the appendix.

At early times after the explosion (around and immediately after the maximum in the optical emission) the gamma rays are absorbed and thermalized by the expanding supernova ejecta. At these times, the luminosity and shape of the optical light curve can be used to constrain the parameters of the magnetar. In this model, the radiation of an input energy reservoir (the spin-down luminosity of a rotating magnetar) diffuses through the ejecta following the analytical solution by \citet{arnett1982} (equation \ref{eq:trapped}). 

The time evolution of the magnetar's spin-down luminosity can be modeled by assuming a rotating dipole magnetic field whose energy loss is dominated by emission of radiation in the gamma-ray and X-ray bands (see Appendix \ref{sec:appendix} for details). 

This luminosity depends on the magnetar initial spin period, surface dipole magnetic field strength, and neutron star mass, $L_{\rm mag}(t, P_0, B, M_{\rm NS})$ (equation \ref{eq:Lmag}). The emitted radiation thermalizes as it diffuses through the ejecta. The conditions of the ejecta determine the optical and gamma-ray outputs, dominated by the values of the ejecta mass, ejecta velocity, and optical and gamma-ray opacities to form $L_{\rm opt}(t, M_{\rm ej}, v_{\rm ej}, \kappa,\kappa_{\gamma})$ (equation \ref{eq:leaking}) and $L_{\gamma}(t, M_{\rm ej}, v_{\rm ej}, \kappa, \kappa_{\gamma})$ (equation \ref{eq:escape}). 

For SN2015bn and SN2017egm, the parameters for the magnetar and the supernova ejecta properties were found by fitting their integrated ultraviolet-optical-infrared (UVOIR) light curves, shown with red points in Figures~\ref{fig:SN2015bn_lc} and \ref{fig:SN2017egm_lc}. All fits were conducted using nonlinear least squares minimization\footnote{\texttt{scipy.optimize.curve\_fit}}.
The best-fit parameters with errors for the magnetar model are given in Table \ref{tab:event_physical_params}. The redshifts and time of peak optical magnitude are shown in the table as listed in The Open Supernova Catalog \citep{Guillochon2016}\footnote{\url{https://sne.space}}.

These parameters are consistent with the results of previous fits \citep{nicholl2018a,nicholl2017c} that took into account both the optical spectral energy distribution and light curve using the open-source code \texttt{MOSFiT} \footnote{\url{https://mosfit.readthedocs.io/en/latest/}}. 
The relative statistical errors on these fit parameters may be optimistic at $\sim10\%$, and the systematic errors will still need to be incorporated for a better understanding the magnetar parameter space. The largest contributor to the magnetar power are the period and magnetic field values, which determine the overall magnitude of the luminosity. The ejecta mass and velocity determine the time to optical peak by the diffusion of the emission through the ejecta.

A particularly important shortfall of this model is the constant effective opacity to both optical and gamma-ray photons, rather than a time-dependent treatment of the opacity. TeV gamma rays interact preferentially with optical photons, so at the time of the peak optical emission, $\gamma\gamma$ absorption by optical photons will be high, reducing any predicted gamma-ray emission by this model. Equation \ref{eq:leaking} is a bolometric luminosity, so it does not take into account the energy and time-dependent opacity, instead fitting a constant effective $\kappa$ and $\kappa_{\gamma}$ to generate the time-dependent optical depth. 

Therefore, Figure~\ref{fig:tau} is used as a guide for when to expect $L_{\gamma}$ to provide an appropriate estimate for the gamma-ray emission. The shaded regions in Figures \ref{fig:SN2015bn_lc} and \ref{fig:SN2017egm_lc} estimate the time periods when photons of the given energies can escape. It is important to reiterate that this model is energy independent, representing the bolometric luminosity not thermalized by the ejecta. This model cannot distinguish the emission between LAT and VERITAS energy bands since it does not consider the physical model of the nebula; the self-consistent model described by \cite{vurm2021} and discussed below will be an attempt to do so explicitly.

Following the methodology in Appendix \ref{sec:appendix} with the magnetar parameters for each SLSN, $L_{\rm mag}(t)$, $L_{\rm opt}(t)$, and $L_{\gamma}(t)$ were calculated and are shown in comparison to the gamma-ray limits in Figures \ref{fig:SN2015bn_lc} and \ref{fig:SN2017egm_lc}.

For SN2015bn (Figure \ref{fig:SN2015bn_lc}), neither the {\it Fermi}-LAT upper limits nor the VERITAS upper limit constrain the predicted escaping luminosity. 
Similarly, for SN2017egm (Figure \ref{fig:SN2017egm_lc}), both the VERITAS and {\it Fermi}-LAT upper limits are not deep enough to constrain the predicted escaping luminosity. An important caveat to these upper limits is that the escaping luminosity may also be emitted at energies not explored here, such as hard X-rays or gamma-rays greater than \SI{ 30}{\TeV}.

The optimal time to observe with a pointed instrument sensitive at a particular photon energy results from a trade-off between the dropping ($\propto t^{-2}$) magnetar luminosity and the rising transparency of the ejecta; predicting the optimal time post-peak to observe requires knowledge of the evolution of the optical spectrum. It is possible to accumulate enough optical data within a few weeks after the optical peak to fit the magnetar model for a reliable prediction of the gamma-ray luminosity. In the case of SN2017egm, the gamma-ray luminosity prediction was anchored by the late optical data points about 1 yr after the explosion. This means that had the VERITAS observations been taken at that point (more than a year earlier than the original observation), they would have been deeply constraining to the magnetar model.

Going beyond these relatively model-independent statements to compare to a more specific spectral energy distribution for the escaping magnetar nebula requires a detailed model for the nebula emission and its transport through the expanding supernova ejecta. Such a model offers preliminary support that a significant fraction of $L_{\gamma}$ may come out in the VHE band \citep{vurm2021}. In this case, the VHE limits on SN2015bn and SN2017egm do not strongly constrain the parameters of the magnetar model, such as the nebular magnetization.

The model of \citet{vurm2021} self-consistently follows the evolution of high-energy electron/positron pairs injected into the nebula by the magnetar wind and their interaction with the broadband radiation and magnetic fields. They found that the thermalization efficiency and the amount of gamma-ray leakage depends strongly on the nebular magnetization, $\varepsilon_B$, i.e. the fraction of residual magnetic energy in the nebula relative to that injected by the magnetar.

The model is simulated for dimensionless $\varepsilon_B$ values set between  $10^{-6}$ and $10^{-2}$; the higher magnetizations lead to greater synchrotron efficiencies, which dominate within a few hundred days, and lead to the optical emission tracking the spin-down luminosity. Lowering the magnetization to $10^{-7}-10^{-6}$ for SLSN-I events like those in this work delays the transition to synchrotron-dominated thermalization, so that the predicted optical emission actually tracks the observed data. 

\begin{figure}[ht]
    \includegraphics[width=1.1\columnwidth]{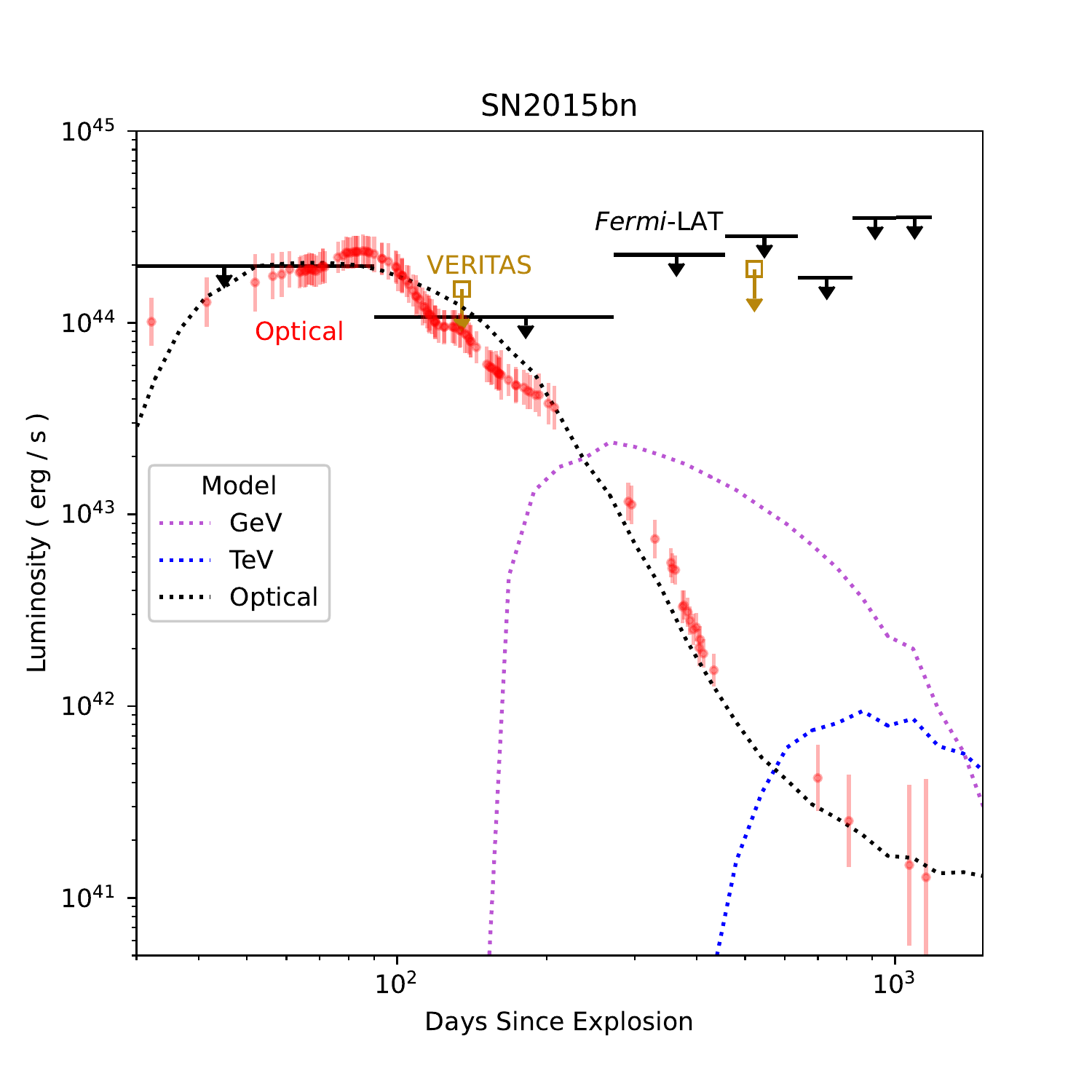}
    \includegraphics[width=1.1\columnwidth]{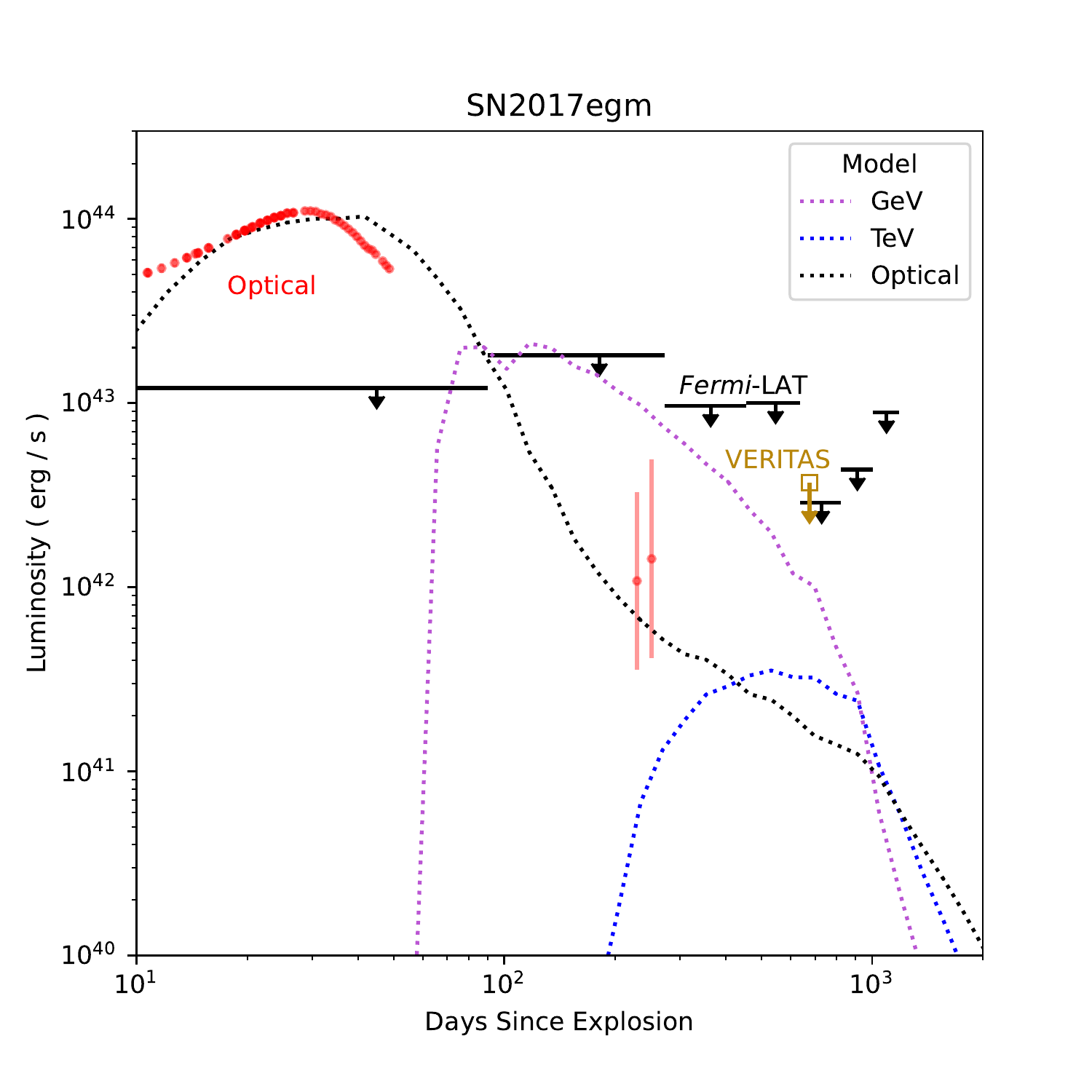}
    \caption{Model light curve for nebular magnetization (from  \cite{vurm2021}) for SN2015bn with $\varepsilon_B = 10^{-7}$ (top panel) and SN2017egm with $\varepsilon_B=10^{-6}$ (bottom panel). }
    \label{fig:Indrek}
\end{figure}

The theoretical light curves and gamma-ray upper limits are shown in Figure ~\ref{fig:Indrek}. \citet{vurm2021} concluded that the predicted low magnetizations constrained by the optical data alone presents new challenges to the theoretical framework regarding the dissipation of the nebular magnetic field. This may invoke magnetic reconnection ahead of the wind-termination shock or near the termination shock through forced reconnection of alternating field stripes described in \citet{komissarov2013}, \citet{lyubarsky2003}, \citet{margalit2018b}. It is also possible that the true luminosity of the central engine decreases faster in time than the simpler $\propto t^{-2}$ magnetic spin down, such that escaping VHE emission is not necessary to explain the model. These VHE upper limits do not rule out this model, and do not settle the challenges inferred by the low magnetization required to fit the optical data. Further observations are needed to probe the nebular magnetization and synchrotron efficiency, and deep VHE observations will contribute to these constraints.

The nondetection of X-rays for both events is consistent with the predictions of \cite{margalit2018a} of a fully ionized ejecta. Even under the most optimistic conditions - an engine that puts 100\% of its spin-down luminosity into ionizing photons of ideal energies - cannot reduce the opacity enough to allow X-rays to escape under the usual assumptions (e.g. spherically symmetric ejecta shell). 

\subsection{Black Hole Central Engine}  \label{sec:BlackHole}

Instead of forming a neutron star like a magnetar, a SLSN-I might form a black hole, in which case the optical peak of the light curve could be powered by energy released from the fallback accretion of ejecta from the explosion (e.g.~\citealt{dexter2013}). Even if a black hole does not form immediately, it could form at late times once the magnetar accretes enough fallback material \citep{moriya2016a}. The main practical difference as compared to a magnetar in section \ref{sec:MagnetarSpinDown} is that the black hole central-engine power would be predicted to decay with the fallback accretion rate $\dot{M}_{\rm fb} \propto t^{-5/3}$ instead of $\propto t^{-2}$. Thus, in principle, for the same luminosity at the time of the optical maximum $t_{\rm pk}$, the central-engine output at times $t \gg t_{\rm pk}$ could be enhanced by a factor $\propto (t/t_{\rm pk})^{1/3} \sim 2$ for $t \sim 1$ yr and $t_{\rm pk} \sim 1$ month, thus tightening our constraints.

In Figures \ref{fig:SN2015bn_lc} and \ref{fig:SN2017egm_lc}, a rough estimate of the maximal engine luminosity in the BH accretion scenario is shown, which is calculated as 

\begin{align}
L_{\mathrm{BH}}=\frac{2^{5/3} L^{\rm pk}_{\rm opt}}{\left(1+\frac{t}{t_{\rm pk}}\right)^{5/3}} \label{eq:L_BH},
\end{align}

where $L^{\rm pk}_{\rm opt}$ is the peak optical luminosity, scaled so that $L_{\rm BH} = L_{\rm opt}$ around the optical peak.

On the other hand, while gamma rays are naturally expected from the ultra-relativistic spin-down-powered nebula of a magnetar, it is less clear that this would be the case for a black hole engine. For instance, the majority of the power from a black hole engine could emerge in a mildly relativistic wind from the black hole accretion disk instead of an ultra-relativistic spin-down-powered pulsar wind.

As seen in both Figure \ref{fig:SN2015bn_lc} (SN2015bn) and Figure \ref{fig:SN2017egm_lc} (SN2017egm), the gamma-ray emission in the black hole scenario is not constrained in the {\it Fermi}-LAT and VERITAS energy bands.

\subsection{Circumstellar Interaction} \label{sec:Circumstellar}

An alternative model for powering the light curve of SLSNe is to invoke the collision of the supernova ejecta with a slower expanding circumstellar shell or disk surrounding the progenitor at the time of the explosion (e.g.~\citealt{smith2006,chevalier2011,Moriya2013a}). Features of this circumstellar model (CSM), such as the narrow hydrogen emission lines that indicate the interaction of a slow-moving gas, provide compelling evidence for this being a powering mechanism for many but not all of the hydrogen-rich class of SLSNe (SLSNe-II; e.g.~\citealt{smith2007,nicholl2020}). 

Shock interaction could in principle also power some hydrogen-poor SLSNe (SLSNe-I), particularly in cases where the circumstellar interaction is more deeply embedded and less directly visible (e.g.~\citealt{sorokina2016,kozyreva2017}). 
There is growing evidence for hydrogen-poor supernovae showing hydrogen features from the interaction in their late-time spectra \citep{Milisavljevic2015,Yan2015,Yan2017,Chen2018,Kuncarayakti2018,Mauerhan2018}. The light echo from iPTF16eh \citep{lunnan2018} implies a significant amount of hydrogen-poor circumstellar medium in a SLSN-I at ${\sim}10^{17}$ cm. However, this material is too distant for the ejecta to reach by the time of maximum optical light and hence cannot be responsible for boosting the peak luminosity.

In principle, the gamma-ray observations of SLSNe can constrain shock models. In many cases, this may not work out since most of the emission from shock-heated plasma is expected to either (1) come out in the X-ray band, as is well studied in other CSM-powered supernovae such as SNe IIn like SN 1998S \citep{Pooley2002}, SN 2006jd \citep{Chandra2012}, SN 2010jl \citep{Chandra2015}, and SNe Ib/c \citep{Chevalier2006} or (2) be absorbed by the surrounding ejecta and reprocessed into the optical band. Thus, these VHE limits on SLSNe do not constrain the bulk of the shock power. 

Higher-energy radiation can be produced if the shocks accelerate a population of nonthermal relativistic particles that interact with ambient ions or the supernova optical emission to generate gamma rays (e.g. via the decay of $\pi^{0}$ generated via hadronic interactions with matter and radiation; e.g., \citealt{murase2011}). However, because shocks typically place a fraction $\epsilon_{\rm rel} \lesssim 0.1$ of their total power into relativistic particles (or even less; \citealt{steinberg2018,fang2019}), the predicted gamma-ray luminosities (matching the same level of optical emission as magnetar models) would be at least 10 times lower than $L_{\gamma}$ predicted by the magnetar nebula scenario, thus rendering our VHE upper limits unconstraining on non-thermal emission from shocks on SN2015bn and SN2017egm. This is consistent with upper limits from the Type IIn SN 2010j from {\it Fermi}-LAT, which \citet{Murase2019} used to constrain $\epsilon_{\rm rel} \lesssim 0.05-0.1$. 

\section{Future Prospects}
\label{sec:future}

These results demonstrate that high-energy gamma-ray observations of SLSN-I are on the brink of enabling constraints on the light curves and even spectral energy distribution of magnetar models. Given the rarity of bright, nearby SLSN-I, and the need to take observations in the optimal window (when $L_{\gamma}$ is near maximum), careful planning will be required to make progress going ahead \citep{Quimby2011,mccrum2015,prajs2017}. The strategy outlined below will focus only on SLSN-I, as type II SLSN are likely to be powered by a mechanism that requires a different consideration of the temporal and spectral evolution of the gamma-ray emission.

Standard arrays of IACTs provide an improved instantaneous sensitivity to gamma-ray emission over {\it Fermi}-LAT due to $10^4$ to $10^5$ larger effective area, counterbalanced in part by the pointed nature of their observations. To propose a strategy, we firstly revisited the characteristics of a large sample of observed SLSNe and performed a systematic study.

\citet{nicholl2017d} fit a sample of 38 SLSNe light curves using MOSFiT to obtain a distribution of magnetar model parameters. This sample is a selection of SLSNe with well-observed events classified as Type I with published data near the optical peak, forming a representative sample of good SLSNe-I for a population study. For each event in this sample, the following was calculated: the escaping gamma-ray luminosity $L_{\gamma}$ following the procedure outlined in Appendix \ref{sec:appendix} and the flux $F_{\gamma} = L_{\gamma}/4\pi D_{\rm L}^{2}$ based on the source luminosity distance $D_{L}$. In performing this analysis, rather than fitting the value of $\kappa_{\gamma}$ individually to each optical light curve (as done in \citealt{nicholl2017d}), the value $\kappa_{\gamma} = \SI{0.01}{cm^{2} g^{-1}}$ is fixed in all events, based on the best-fit to SN2015bn (given its particularly high-quality late-time data, which provides the most leverage on $\kappa_{\gamma}$). 

\begin{figure}[ht!]
    \centering
    \includegraphics[width=\columnwidth]{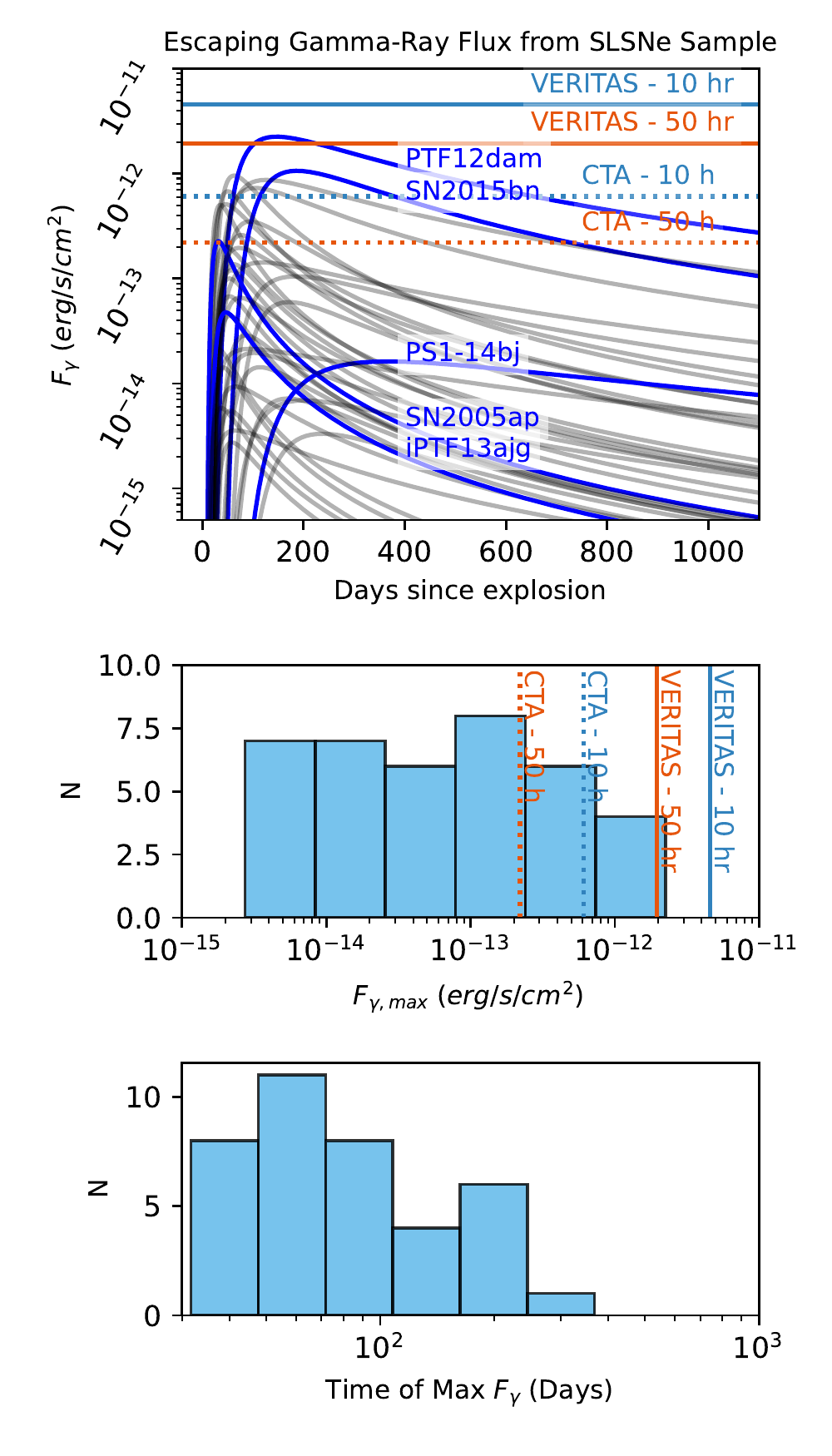}
    \caption{Top: escaping gamma-ray luminosity $L_{\gamma}(t)$ for the sample of SLSNe fit by \citet{nicholl2017d}. Five well-studied SN are highlighted in blue, including SN2015bn. Overplotted are the VERITAS and CTA sensitivity curves for various exposures.  Middle: distribution of peak escaping gamma-ray flux $F_{\rm \gamma,max} = {\rm max}[L_{\gamma}]/4\pi D^{2}$, for the light curves from the top panel where $D$ is the distance to each source. Again, VERITAS and CTA sensitivities for different exposures are shown as vertical dashed lines. Bottom: distributions of times since explosion to reach the maximum gamma-ray flux $F_{\gamma,max}$ from $F_{\gamma}$ above.} 
    \label{fig:distributions}
\end{figure}

The results for $F_{\gamma}(t)$ are shown in the top panel of Figure~\ref{fig:distributions}. In the magnetar model, the predicted gamma-ray flux could emerge anywhere across the HE to VHE bands; hence, it represents an upper limit on flux in the bands accessible to {\it Fermi}-LAT and IACTs. The bottom two panels of Figure~\ref{fig:distributions} show the distribution of the peak escaping flux $F_{\rm \gamma, max}$ and time of the peak flux relative to the explosion. For most SLSNe-I presented here, $F_{\rm \gamma, max}$ is well below the sensitivity of VERITAS and even the future Cherenkov Telescope Array (CTA) \citep{thectaconsortium2019}. Also note that the characteristic timescale to achieve the peak gamma-ray flux is $\approx 2-3$ months from the explosion. This timescale occurs approximately at the same time as when the optical depth of the ejecta to VHE emission falls below unity, when the VHE photons can escape (Figure~\ref{fig:tau}).  

\begin{figure*}[ht]
    \centering
    \includegraphics[width=.95\textwidth]{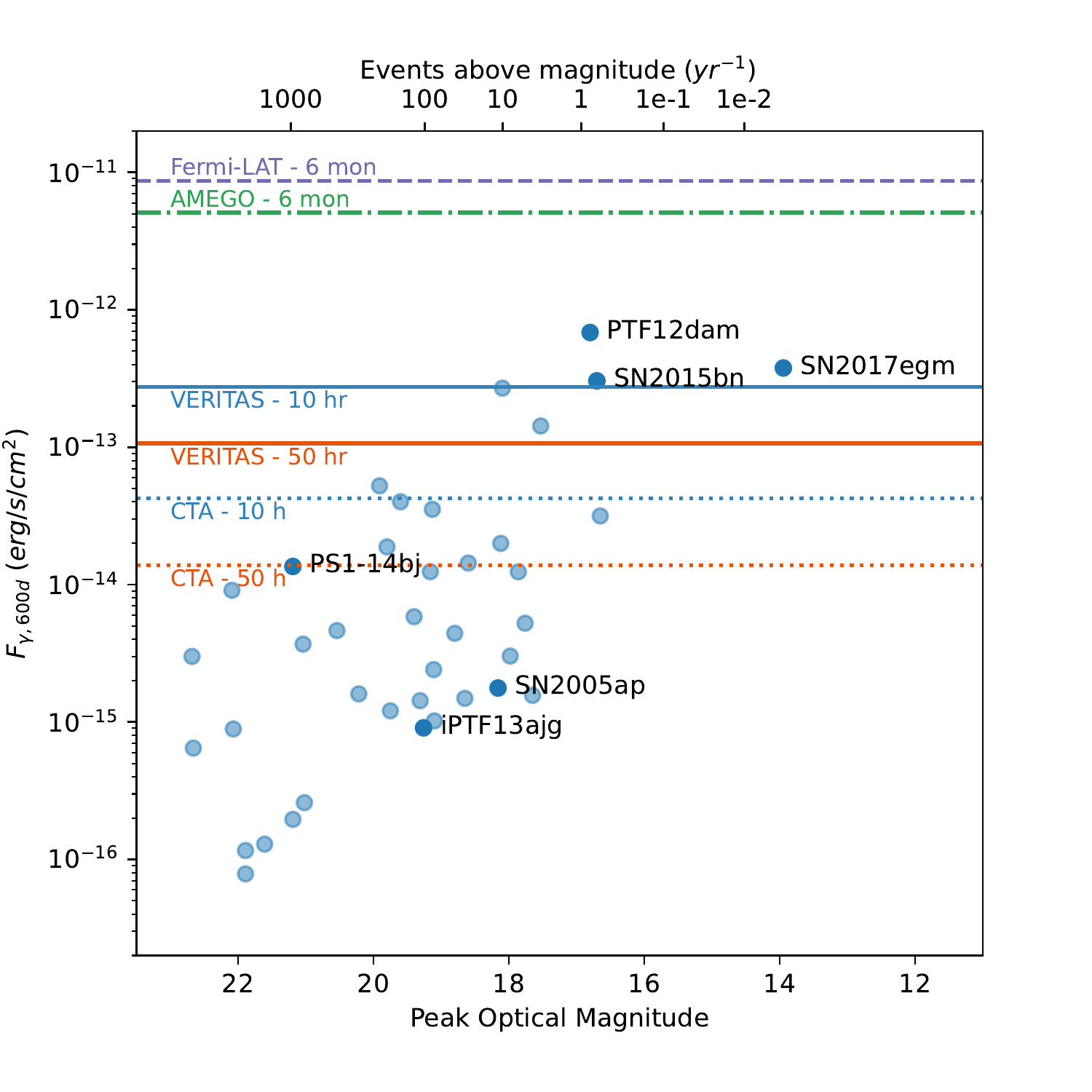}
    \caption{
    Blue dots show the peak optical apparent magnitudes of a sample of SLSNe-I \citep{nicholl2017d} as a function of their predicted maximum gamma-ray luminosity at 600 days after explosion ($F_{\gamma, 600d}$). The top axis shows the approximate rate of events above the given peak optical magnitude, calculated using the method described in the main text. Peak maximum gamma-ray luminosities are calculated from fits of optical data with fixed $\kappa_{\gamma} = 0.01$ cm$^{2}$ g$^{-1}$. Integral sensitivities of various instruments are overplotted for different exposures. Solid lines: VERITAS 10 and 50 hr integral sensitivities above 220 GeV. Dotted lines: CTA (in development) 10 and 50 hr integral sensitivities above 125 GeV as estimated from 50 hr Monte Carlo simulations of the southern array \citep{thectaconsortium2019} and extrapolated to 10 hours. Similar extrapolation is done for {\it Fermi}-LAT from 10 yr to 6 months \citep{nolan2012} (dashed line). Proposed project AMEGO integral sensitivity above 100 MeV for 6 month observation window is also plotted (dashed-dotted line) \citep{kierans2020}. 
    }
    \label{fig:Optical_to_Escaping_Fluxes}
\end{figure*}

Figure~\ref{fig:Optical_to_Escaping_Fluxes} shows $F_{\rm \gamma,600 d}$ as a function of the peak optical magnitude of the SLSNe-I from the same sample as in Figure~\ref{fig:distributions}. The selection of fluxes at \SI{600}{\day} approximates the time when the effective opacity to 1 TeV photons reaches 1, based on Figure~\ref{fig:tau}. The top axis also gives the all-sky rate of SLSNe-I above a given peak optical magnitude, which is estimated using the magnitude distribution of SLSNe-I and assuming they occur at a comoving volumetric rate of $ R(z)=19(1+z)^{3.28}\SI{}{\, Gpc^{-3}\, yr^{-1}}$ following \citet{nicholl2017b,lunnan2018,decia2018}. This estimation captures the general volumetric rate of events, but is unreliable for exceptionally bright events such as SN2017egm due to the small population for estimating the magnitude normalization. A bright event like SN2017egm may actually happen more often than once a century.

Shown for comparison in Figure~\ref{fig:Optical_to_Escaping_Fluxes} are the integral sensitivities of various gamma-ray instruments for different exposures. For IACT instruments such as VERITAS and the future CTA, sensitivity is defined as the minimum flux necessary to reach $5 \sigma$ detection of a point-like source, requiring at least 10 excess gamma rays and the number of signal counts at least $5\%$ of the number of background counts. For VERITAS, the sensitivity was calculated using observed Crab Nebula data to estimate the rates of signal and background photons with cuts optimized for a $\Gamma = -2.5$ power-law spectrum, and then rescaled for the appropriate observation time \citep{Park2015}. For CTA, Monte Carlo simulations were used to derive angular resolution, background rates and energy dispersion features -- the instrument response functions (IRF) -- based on the Prod3b-v2 telescope configuration for the Southern site and its atmosphere \citep{cherenkovtelescopearrayobservatory2016}. These IRFs are publicly available and were analyzed using the open-source CTOOLS\footnote{\url{http://cta.irap.omp.eu/ctools/}} \citep{Knodlseder2016}. A power-law spectral model was used to estimate the integral sensitivity above \SIlist{0.125;1}{\TeV} each for observations of \SIlist{10;50}{\hour} (see \citealt{Fioretti2016} for further discussion on CTA integral sensitivity).

Based on this systematic study, we propose the following observation strategy: (1) Receive automated public alert and Type I classification of SLSN from a survey instrument such as the Zwicky Transient Facility (ZTF). Classification is generally determined by identification of early spectral components such as $O_{II}$ absorption features. (2) During the multiday rise and fall of bolometric optical light curve, fit the magnetar model ($L_{\rm opt}$, yielding parameters for $L_{\rm mag}$ and $L_{\gamma}$) (3) Compare $L_{\gamma}$ to the telescope sensitivity at the appropriate day when the effective $\gamma$-$\gamma$ opacity falls below ${\sim1}$ for the telescope's sensitive energy range (see Figure~\ref{fig:tau}). In the case of IACTs sensitive to energies above $\SI{100}{\GeV}$, the gamma rays will escape the magnetar a few hundred days after explosion, requiring a bright SLSN-I that will power gamma rays for as much as two years. 
%% reorder ????? ^^^

Estimating ${\sim}35\%$ of all-sky visibility at VERITAS due to Sun, Moon, and seasonal weather cut, and above 60\textdegree~ elevation, VERITAS is capable of detecting up to ${\sim}0.4$ and ${\sim}4$ SLSNe-I per year for \SI{10}{h} and \SI{50}{h} exposures, respectively. The next-generation CTA observatory will be able to detect as many as ${\sim}8$ and ${\sim}80$ events for \SI{10}{hr} and \SI{50}{h}, respectively, assuming a larger sky visibility fraction of ${\sim}80\%$ when both North and South arrays are included. On the other hand, SLSNe at greater distances also imply a stronger role of $\gamma-\gamma$ interactions on the EBL in suppressing the $\gtrsim$ TeV emission, decreasing the observed integral flux by as much as 60 times at redshifts near 0.5 in the VERITAS energy range. 

\begin{figure}[ht]
    \centering
    \includegraphics[width=\columnwidth]{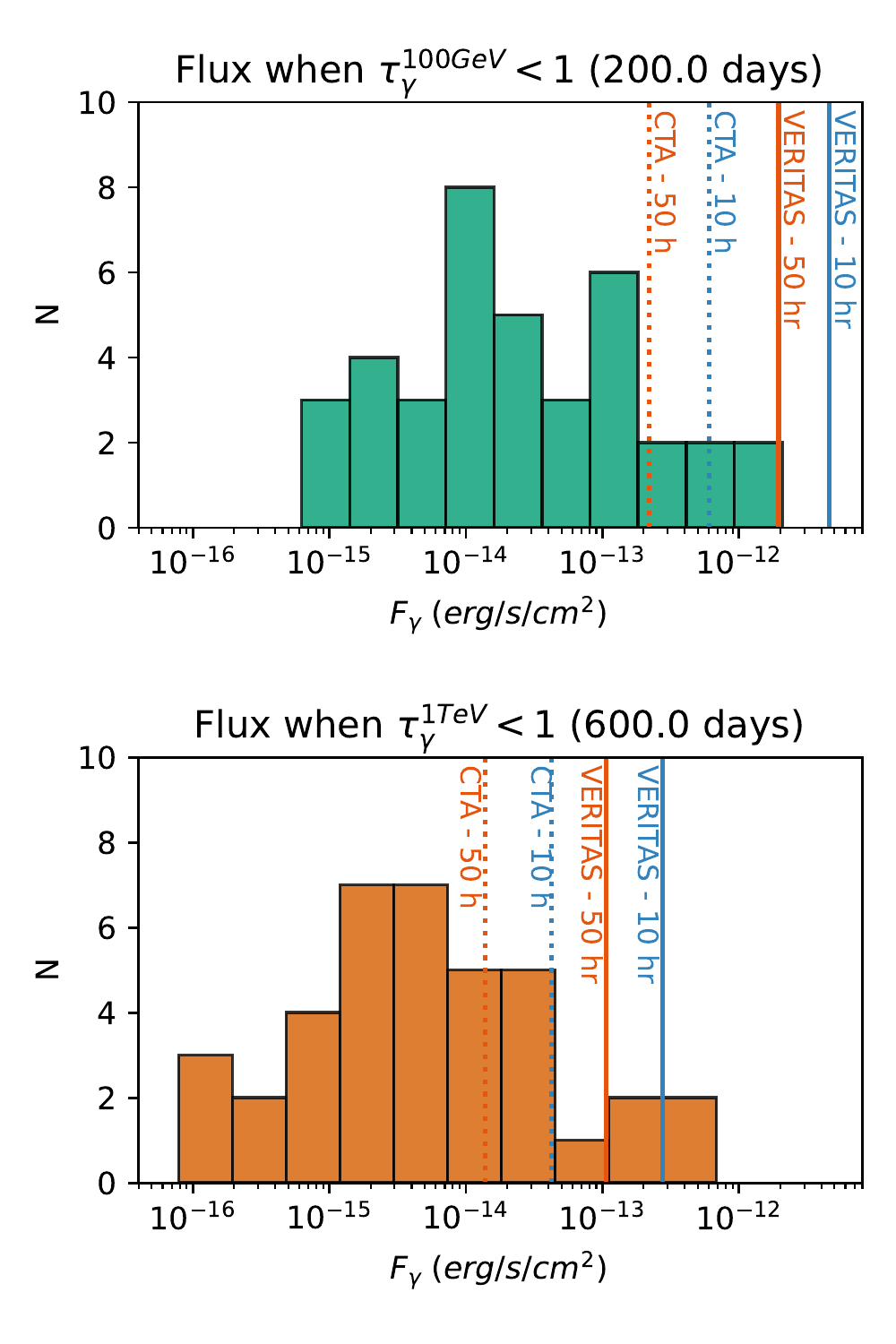}
    \caption{ Distribution of gamma-ray luminosities $L_{\gamma}$ at $t=\SI{200}{\day}$ (top) and $t=\SI{600}{\day}$ (bottom), when the optical depth for \SI{100}{\GeV} and \SI{1}{\TeV} photons drops below 1, calculated for a sample of 38 SLSNe \citep{nicholl2017d}}
    \label{fig:TeV_gamma_flux_sens}
\end{figure}

Figure~\ref{fig:TeV_gamma_flux_sens} shows the distribution of fluxes at \SI{200}{\day} and \SI{600}{\day} which are approximate average dates when the opacity to \SI{100}{\GeV} and \SI{1}{\TeV} photons falls below 1, respectively, and they are able to escape the ejecta. Accounting for this time delay for the opacity to drop, the expected rate of bright events drops by another 3 to 15 times. While past observations have not been followed up until this publication, the distribution of predicted gamma-ray fluxes hints that, particularly for \SI{100}{\GeV} photons, future SLSN-I will be observable with current and planned observatories.

\section{Conclusion}
\label{sec:conclusions}
SLSN-I are potential gamma-ray emitters, and this paper provides the first upper limits at different times after the optical outburst for two good candidates. The reported upper limits approach the magnetar spin-down luminosity limit of SN2015bn and SN2017egm. While the expected gamma-ray luminosity in either the magnetar central-engine scenario or the shock-acceleration scenario is not constrained by these limits, a relativistic jet powered by fallback accretion onto a black hole is disfavored in both cases. We explore prospects for obtaining improved VHE gamma-ray constraints in the future by current and planned IACTs. We estimate the Type I SLSNe rate for VERITAS and CTA, considering observation constraints and the time delay due to the optical depth. For sufficiently nearby and bright SLSN-I, 0.4 and 4 events per year can be observed by VERITAS from 10 hr and 50 hr observation, respectively, and similarly rates of 8 and 80 events per year can be expected by CTA.

\acknowledgments 
{This research is supported by grants from the U.S. National Science Foundation and the Smithsonian Institution, by NSERC in Canada, and by the Helmholtz Association in Germany. M.N. is supported by the European Research Council (ERC) under the European Union’s Horizon 2020 research and innovation program (grant agreement No.~948381) and by a Fellowship from the Alan Turing Institute. I.V. acknowledges support by the ETAg grant PRG1006 and by EU through the ERDF CoE grant TK133. V.V.D.'s work is supported by NSF grant 1911061 awarded to the University of Chicago (PI: Vikram Dwarkadas). We acknowledge the excellent work of the technical support staff at the Fred Lawrence Whipple Observatory and at the collaborating institutions in the construction and operation of the instrument. 
This research has made use of the CTA instrument response functions provided by the CTA Consortium and Observatory, see \url{http://www.cta-observatory.org/science/cta-performance/} (version prod3b-v2) for more details.
}

\software{fermipy (v0.19), \citep{wood2017},
astropy \citep{Robitaille2013,Price-Whelan2018},
CTOOLs, \citep{Knodlseder2016a}, superbol \citep{nicholl2018} , EventDisplay \citep{Maier2017}}, VEGAS \citep{cogan2008}

\facilities{VERITAS, {\it Fermi}-LAT}

\clearpage

\appendix

\section{Magnetar Light Curve Model}
\label{sec:appendix}

Following \citet{woosley2010} and \citet{kasen2010}, the spin-down power of a strongly magnetized, young neutron star (``magnetar") at a time $t$ after its birth is given by the magnetic dipole luminosity,

\begin{align}
    L_{\rm mag}(t) = \frac{E_{\rm rot}}{\tau_p} \frac{2}{(1+2t/\tau_p)^2}.
    \label{eq:Lmag}
\end{align}

Here, $E_{\rm rot} = (1/2)I_{\rm NS} \Omega^2_i \approx 2.6\times10^{52} (M_{\rm NS}/1.4M_{\odot})^{3/2} P^{-2}$ erg is the magnetar rotational energy, $I_{\rm NS}$ is its moment of inertia, and $M_{\rm NS}$ is the neutron star mass. The spin-down time is given by $\tau_p = (6I_{\rm NS}c^3/B^2 R^6_{\rm NS}\Omega^2_{i}) \approx 1.3\times10^5 B^{-2}_{14} P^2 \left(M_{\rm NS}/1.4M_{\odot}\right)^{3/2} \sin^{-2}{\left(\theta_{B_{\perp}}=90\degree\right)}$ s, where $B = 10^{14}B_{14}$ G is the dipole magnetic field strength, $P$ (ms) is the birth spin period ($\Omega = 2\pi/P$ is the birth angular velocity), and $\theta_{B_{\perp}}$ is the inclination angle of the magnetic dipole axis relative to the rotation axis.

Magnetar energy deposited behind the ejecta shell is assumed to thermalize and then diffuse outward through the ejecta as electromagnetic radiation. This occurs over a characteristic diffusion time \citep{arnett1982}

\begin{align}
\tau_m = \left(\frac{2\kappa M_{\rm ej}}{\beta c v}\right)^{1/2},
\end{align}
where $M_{\rm ej}$, $v$, and $\kappa$ are the total mass, mean velocity, and (optical wavelength) opacity of the supernova ejecta, respectively, and $\beta \approx 13.7$ is a constant. In most cases, $\tau_{\rm m}$ sets the peak timescale of the supernova light curve. 

To allow for the possibility that high-energy photons from the central magnetar nebula can escape instead of thermalizing, one can apply a trapping coefficient $(1 - e^{-\tau_{\gamma}})$ where the optical depth of the ejecta to gamma-rays can be written as $\tau_{\gamma} = At^{-2}$, where $A\equiv  (3\kappa_{\gamma}M_{\rm ej}/4 \pi v_{\rm ej}^{2})$ and $\kappa_{\gamma}$ is an effective gamma-ray opacity \citep{clocchiatti1997,chatzopoulos2012a}. For large $\tau_{\gamma} \gg 1$ (early times), the effects of gamma-ray leakage are small and the optical light curve (after the optical peak, at times $\gtrsim \tau_{\rm m}$) will follow the spin-down luminosity, i.e. $L_{\rm opt} \approx L_{\rm mag}$. However, at late times when $\tau_{\gamma} \ll 1$, one has $L_{\rm opt} \ll L_{\rm mag}$, with the remaining luminosity $L_{\gamma} = L_{\rm tot}-L_{\rm opt}$ escaping as gamma rays.

More precisely, the luminosity of the magnetar (eq.~\ref{eq:Lmag}) that escapes the ejecta via photon diffusion by time $t$ is calculated by \citep{arnett1982,inserra2013}

\begin{align}
    L_{\rm tot}(t) = \frac{2}{\tau_m}e^{-\left(\frac{t}{\tau_m}\right)^2} \times \int_0^{t}\text{d}t^{\prime} L_{\rm mag}(t^{\prime})e^{\left(\frac{t^{\prime}}{\tau_m}\right)^2} \frac{t^{\prime}}{\tau_m}\label{eq:trapped}
\end{align}

Of this total luminosity, only a fraction is able to thermalize and hence power the optical supernova light curve,

\begin{align}
    L_{\rm opt}(t) = (1 - e^{-At^{-2}}) L_{\rm tot}(t), \label{eq:leaking}
\end{align}

with the remainder instead escaping as gamma-rays,

\begin{align}
L_{\gamma} = L_{\rm tot}-L_{\rm opt}
 = e^{-At^{-2}} \frac{2 E_p}{\tau_{\rm p}\tau_m^2}e^{-(\frac{t}{\tau_{\rm m}})^2} \int_0^{t}\text{d}t^{\prime} \frac{2}{(1+2t^{\prime}/\tau_{\rm p})^2}e^{\left(\frac{t^{\prime}}{\tau_{\rm m}}\right)^2} t^{\prime}\label{eq:escape}
\end{align}

\bibliographystyle{aasjournal}
\bibliography{references}

\end{document}